\documentclass[twocolumn]{aastex62}
\usepackage{amsmath,amstext}
\usepackage[T1]{fontenc}
\usepackage{natbib}
\usepackage{sidecap}

\begin{document}

\title{Uranus's influence on Neptune's exterior mean motion resonances} 
\shorttitle{}

\author[0009-0000-2923-9953]{Severance Graham}
\correspondingauthor{Severance Graham}
\email{sevgraham77@arizona.edu}
\affiliation{Lunar and Planetary Laboratory, The University of Arizona, 1629 E University Blvd, Tucson, AZ 85721}

\author[0000-0001-8736-236X]{Kathryn Volk}
\affiliation{Planetary Science Institute, 1700 East Fort Lowell, Suite 106, Tucson, AZ 85719, USA}
\affiliation{Lunar and Planetary Laboratory, The University of Arizona, 1629 E University Blvd, Tucson, AZ 85721}

\begin{abstract}
Neptune's external mean motion resonances play an important role in sculpting the observed population of transneptunian objects (TNOs). 
The population of scattering TNOs are known to `stick' to Neptune’s resonances while evolving in semimajor axis ($a$), though simulations show that resonance sticking is less prevalent at $a\gtrsim200-250$~au.
Here we present an extensive numerical exploration of the strengths of Neptune’s resonances for scattering TNOs with perihelion distances $q=33$~au.
We show that the drop-off in resonance sticking for the large $a$ scattering TNOs is not a generic feature of scattering dynamics, but can instead be attributed to the specific configuration of Neptune and Uranus in our solar system.
In simulations with just Uranus removed from the giant planet system, Neptune’s resonances are strong in the scattering population out to at least $\sim300$~au.
Uranus and Neptune are near a 2:1 period ratio, and the variations in Neptune’s orbit resulting from this near resonance are responsible for destabilizing Neptune’s resonances for high-$e$ TNO orbits beyond the $\sim20$:1 resonance at $a\approx220$~au. 
Direct interactions between Uranus and the scattering population are responsible for slightly weakening Neptune’s closer-in resonances.
In simulations where Neptune and Uranus are placed in their mutual 2:1 resonance, we see almost no stable libration of scattering particles in Neptune’s external resonances.
Our results have important implications for how the strengths of Neptune’s distant resonances varied during the epoch of planet migration when the Neptune-Uranus period ratio was evolving. 
These strength variations likely affected the distant {scattering}, resonant, and detached TNO populations.
\end{abstract}

\section{Introduction} \label{sec:intro}

Many known trans-neptunian objects (TNOs) are observed to orbit within Neptune’s external mean-motion resonances (see, e.g., review by \citealt{Gladman:2021}; we refer to mean-motion resonances simply as resonances hereafter). 
In particular, the observed set of TNOs on orbits with semimajor axes beyond the classical belt region, $a\gtrsim50$~au, imply the existence of large intrinsic populations in many of Neptune's distant resonances (see \citealt{Crompvoets:2022} for recent population estimates). 
TNOs in Neptune's resonances can either be objects captured onto resonant orbits that are stable for a gigayear or longer during the epoch of giant planet migration, or they can be objects that have temporarily `stuck' to resonances while evolving in semimajor axis due to perturbations from the giant planets (see, e.g., review by \citealt{Malhotra:2019}).
In either case, to understand the significance of the observed resonant TNO populations, we need improved models of {both the emplacement of dynamically excited TNOs onto their present day orbits (see, e.g., recent works by \citealt{Huang:2022,Nesvorny:2023,Bottke:2023,Kaib:2024} and reviews by \citealt{Morbidelli:2020,Gladman:2021}) as well as} the dynamical extent of Neptune’s resonances in the distant solar system.
\cite{Volk:2022} recently showed that, in the current solar system, Neptune's resonances remain strong out to surprisingly large semimajor axes of several hundred au in the high-perihelion population of distant TNOs.
Here we focus on the role of distant resonances in the lower-perihelion scattering TNO population where the phenomenon of resonance sticking is known to occur \citep[e.g.][]{Gallardo:2006KZ,Lykawka:2007,Yu:2018}.

Scattering TNOs are typically defined as objects whose semimajor axes experience significant changes over short dynamical timescales due to relatively close approaches to Neptune at perihelion; the widely used \cite{Gladman:2008} scheme defines them as objects that experience $\Delta a>1.5$~au over a 10 Myr integration of their present-day orbit. 
Objects with unchanging, non-resonant semimajor axes beyond the classical belt region (the classical belt being typically defined as sunward of Neptune's 2:1 resonance) belong to the dynamically `detached' TNO population.
The perihelion distance ($q$) boundary between detached and scattering TNOs is complex and semimajor axis dependent (see, e.g., discussions in \citealt{Saillenfest:2020} and \citealt{Batygin:2021}), but non-resonant TNOs with $q\lesssim37$~au are quite likely to experience significant changes in orbital energy, and therefore changes in $a$, as they evolve over time.
This mobility in $a$ allows the particles to encounter different resonances, and they often experience repeated bouts of temporary libration in these resonances in between scattering events. 
\cite{Yu:2018} examined the detailed sticking behavior of scattering TNOs with $a<100$~au and found that at any given time, $\sim40\%$ of the scattering population is temporarily librating in one of Neptune's many resonances; these temporary resonance sticks spanned anywhere from a few libration cycles ($\sim10^5$ years) to a gigayear.
{The wide range of sticking timescales, as well as the large number of potential resonances with Neptune to stick to, makes resonance sticking a challenging phenomena to quantify. 
In this work, we focus on the short-term behavior of particles in the strongest of Neptune's external resonances to better understand trends in resonance sticking behavior.}

A reasonable lower boundary in $q$ for the population of scattering TNOs most likely to experience significant  resonance sticking is $q\approx33$~au. 
Scattering TNOs with lower $q$ experience very strong encounters with Neptune (and the other giant planets for very low-$q$ TNOs) that shorten their dynamical lifetimes compared to the $q\ge33$~au TNOs \citep[see, e.g.,][]{Tiscareno:2003}, reducing the time available for resonance sticking.
{Most of our simulations discussed in the following sections will focus on this $q=33$~au scattering population that we expect to be quite sensitive to changes in Neptune's resonance strengths.
We note that we quantify the strengths of Neptune's resonances based on how long particles tend to remain in or stick to each resonance in numerical simulations {(following, e.g., \citealt{Lykawka:2007})}; as we discuss below, this is a more relevant metric  for scattering TNOs than the usual analytically calculated disturbing function derived strengths.}

\cite{Lykawka:2007} performed numerical simulations of the scattering population with the current giant planets and examined the frequency and locations of resonance sticking events. 
They found that sticking was unlikely for scattering particles with semimajor axes $a\gtrsim250$~au. 
Presumably this drop-off in resonance sticking is due to the resonances becoming weak at these semimajor axes compared to the scattering perturbations. 
{Such a drop-off in resonance strengths at large period ratios due to scattering perturbations has been shown for comets interacting with Jupiter's exterior N:1 resonances based both on analytical modeling \citep{Chambers:1997} and more complete numerical integrations \citep{Fernandez:2016}.}
Analytical models \citep[e.g.][]{Pan:2004,Batygin:2021} of Neptune's resonances predict that the onset of resonance overlap, and thus the destruction of stable {and quasi stable resonant} libration zones, for perihelion distances similar to those of scattering TNOs should occur at semimajor axes {$a\approx500$~au for Neptune's N:1 resonances (see equation 30 in \citealt{Pan:2004}).
We would expect to see resonance sticking continue to occur out to these semimajor axes and then stop when the resonant islands are no longer distinct or stable enough for even temporary libration.  
In contrast, resonance sticking in full numerical simulations of the scattering population drops off much closer in at $a\approx250$~au \citep{Lykawka:2007}.}
This mismatch between the scattering simulation results and the analytical resonance model expectations {for Neptune's resonances is likely} partly due to the inherent limitations of analytical approaches in terms of the number of resonances modeled; typically analytical models only consider a subset of the lowest-order resonances, which in the case of the scattering population are the N:1 and N:2 resonances.
{While these simplified analytical models appear to work well for predicting the behavior of Jupiter's resonances \citep[e.g.][]{Chambers:1997}, they are not sufficient in the case of Neptune's resonances.}
{The higher-order resonances surrounding Neptune's N:1 and N:2 resonances help weaken these resonances, shrinking their libration zones at large $a$ and large $e$.
This leads to faster scattering timescales for low-$q$ TNOs and less resonance sticking overall.}
Recent work by \cite{Hadden:2024} demonstrates how numerical mapping approaches, which do not suffer from {resonance order limitations}, can be used to model resonant and scattering dynamics more accurately in the one-planet case.

However, we will show in this work that there is an additional reason for the mismatch between models of Neptune's resonances {in the one-planet case} and the findings of full scattering simulations: dynamical interactions between Uranus and Neptune strongly influence the strength of Neptune's resonances and drive the $a\approx250$~au drop-off in resonance sticking in the present-day scattering population.
In the low-perihelion regime of the scattering TNOs, single-planet models (whether analytical or numerical) fail to capture this dynamical influence that weakens Neptune's resonances.
Resonance sticking is a very important dynamical process in shaping the present-day scattering TNO population, but variations in the strength of Neptune's resonances during the epoch of planet migration would also leave an imprint on the {production of the initial population of scattering, resonant and detached TNOs}.  
It is thus important to have more complete models of Neptune's external resonances and a better understanding of what affects their strengths.

The rest of the paper is organized as follows. 
In Section~\ref{sec:methods} we describe how we numerically model Neptune's resonances, show how we identified Uranus's important role in affecting them, and describe how we quantify resonance strengths from our simulations.
In Section~\ref{sec:NUperiod} we present our exploration of how the Neptune-Uranus period ratio affects Neptune's external resonances, and we isolate the specific dynamical effects responsible for destabilizing the distant resonances in today's scattering population.
Section~\ref{sec:discussion} discusses the implications of our findings for the scattering population and for using the distant resonant and detached populations to constrain the migration history of the giant planets.
We summarize our results in Section~\ref{sec:summary}.

\section{Simulating Neptune's resonances} \label{sec:methods}

\begin{figure}
    \centering
    \begin{tabular}{c}
    \includegraphics[width=2.15in]{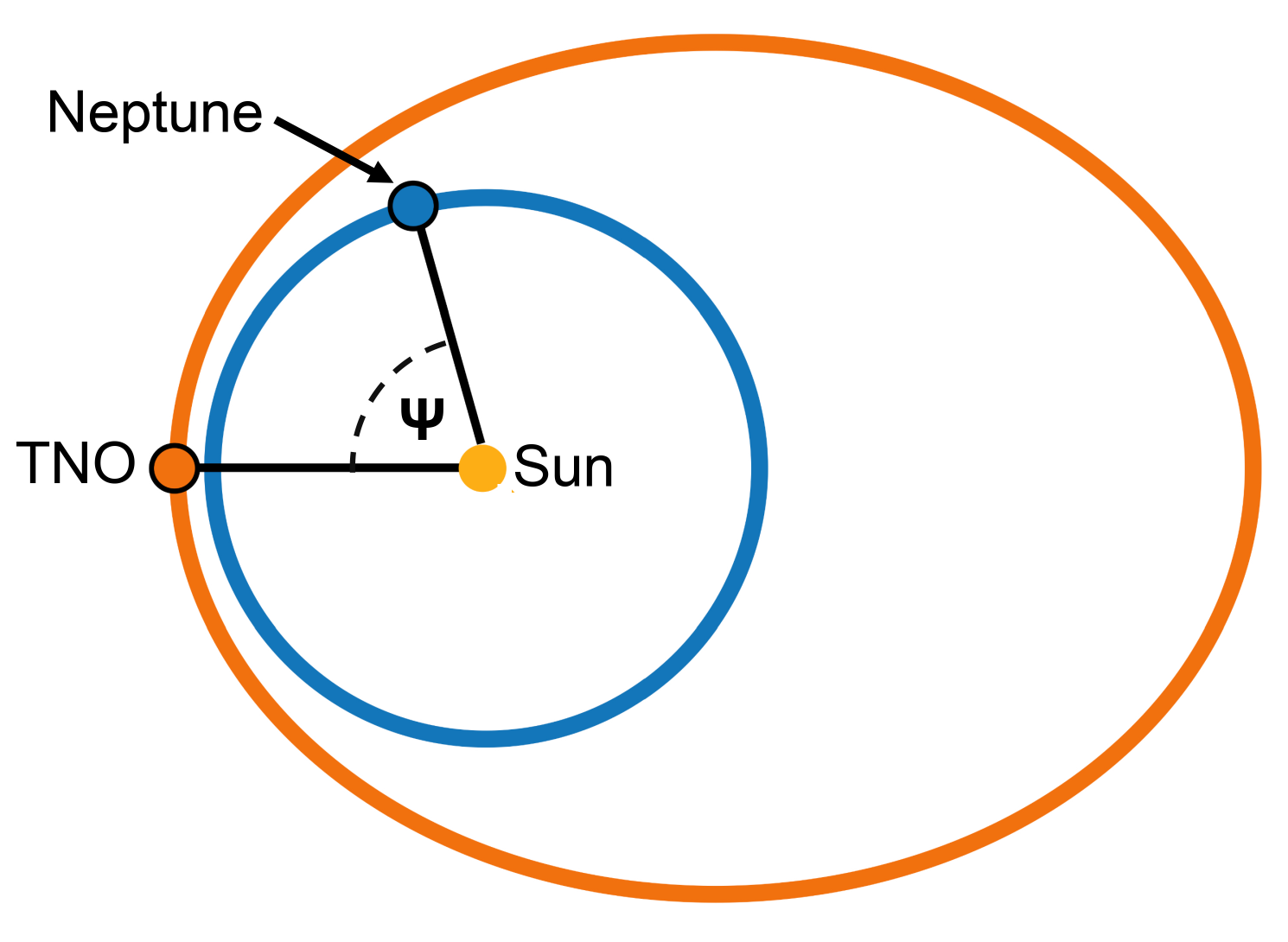}\\
    \includegraphics[width=2.95in]{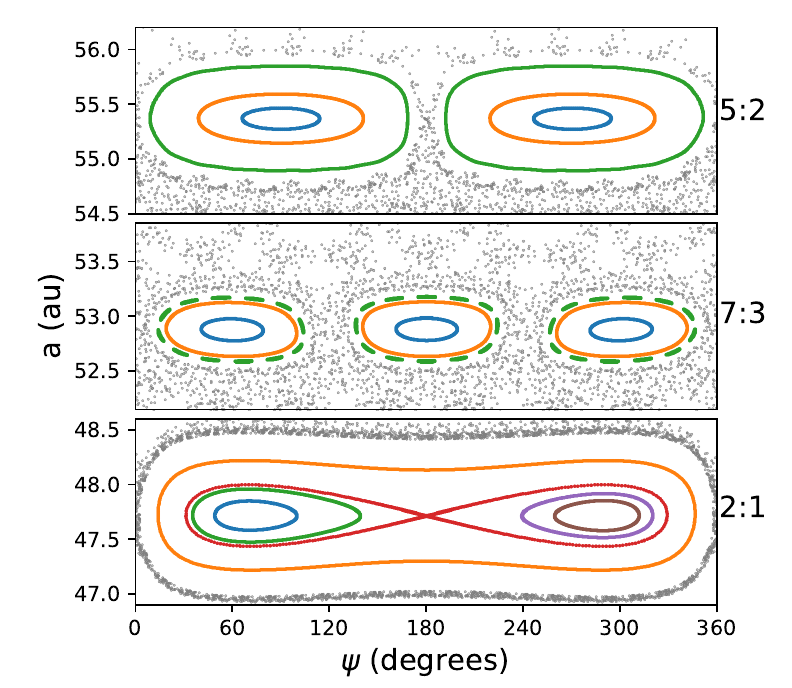}
    \end{tabular}
    \caption{Top: Illustration of the angle $\psi$, which describes a TNO's position relative to Neptune when the TNO is at perihelion.
    Bottom: Example evolution of resonant (colored) and non-resonant (gray) particles with $q=33$~au in $a$-$\psi$ space for the 2:1, 7:3, and 5:2 resonances in the circular restricted three-body problem (Sun + Neptune + massless test particles).
    Within each panel, individual resonant particles are given distinct colors; note the clear islands traced out by the resonant particles.}
    \label{fig:psi}
\end{figure}

A conceptually simple way to explore the strength and boundaries of Neptune’s resonances is to produce Poincar\`e maps, also called surfaces of section. 
In a simplified system with just the Sun and Neptune present, these maps of any particular resonance can be produced by simulating test particles with fixed perihelion distances over a small range of semimajor axis, $a$, near the exact resonant value and recording the evolution of these particles’ orbits and the angle $\psi$ between the particle and Neptune at every perihelion passage (following, e.g., \citealt{Wang:2017,Malhotra:2018,Lan:2019}); $\psi$ is illustrated in the top panel of Figure~\ref{fig:psi}. 
{We note that here and throughout the entire manuscript, we use barycentric rather than heliocentric semimajor axes.}
For particles with perihelion distances consistent with scattering TNOs ($q\approx33-37$~au), those on resonant orbits remain relatively stable in semimajor axis, tracing out distinct resonant islands in $a$-$\psi$ space, while those on scattering orbits wander randomly through that space filling in a sea of background points. 

Figure~\ref{fig:psi} shows example $a$-$\psi$ evolution for $q=33$~au particles in and near the 2:1, 7:3, and 5:2 resonances in the circular restricted three-body problem with the Sun and Neptune as perturbers. 
The details of our methods and how these maps are generated are described in Section~\ref{ss:exploration}, but we briefly highlight here a few key features of Neptune's external resonances.
In general, a particle in a $p$:$q$ external resonance will visit $q$ distinct resonant islands over a resonant cycle {($q$ orbital periods of the resonant particle)}; so N:2 resonant particles (such as the 5:2 particles in Figure~\ref{fig:psi}) trace out two islands, N:3 particles trace out three islands, and so on.
Particles in external N:1 resonances follow this pattern in that individual particles trace out a single island, but the structure of these resonances is distinct in that there are three possible libration islands to choose from. 
These are the two so-called asymmetric islands centered near $\psi\sim90^\circ$ and $\psi\sim270^\circ$\footnote{{The exact center of libration for the asymmetric islands of an N:1 resonance are eccentricity-dependent (see, e.g., \citealt{Nesvorny:2001}).}} (note in Figure~\ref{fig:psi} that these two islands are traced by distinct, differently colored particles) and the larger symmetric island centered on $\psi=180^\circ$. 
We refer the reader to \cite{Volk:2022} for additional details on the structure of these resonances.
In our later maps of Neptune's resonances in various simulations, we examine how all of these resonant islands become stronger, weaker, or disappear entirely.

In Section~\ref{ss:exploration}, we describe how the simplified three-body model shows that Neptune’s resonances should remain strong out to semimajor axes of several hundred au in the scattering population.
We then expand our simulations to map Neptune's resonances under the influence of the other giant planets (following the methods described in \citealt{Volk:2022}) to show that the distant resonances become much weaker for scattering TNOs. 
In Section~\ref{ss:destabilization} we show that Uranus is primarily responsible for this weakening.
In Section~\ref{ss:strengths} we describe how we quantify resonance strengths in our simulations and present the present-day strengths of Neptune's resonances in the scattering population.

\subsection{Exploring resonances in the scattering TNO population}\label{ss:exploration}

\begin{figure*}
    \centering
    \begin{tabular}{c c}
    Neptune only  & all four giant planets \\ 
        \includegraphics[width=2.75in]{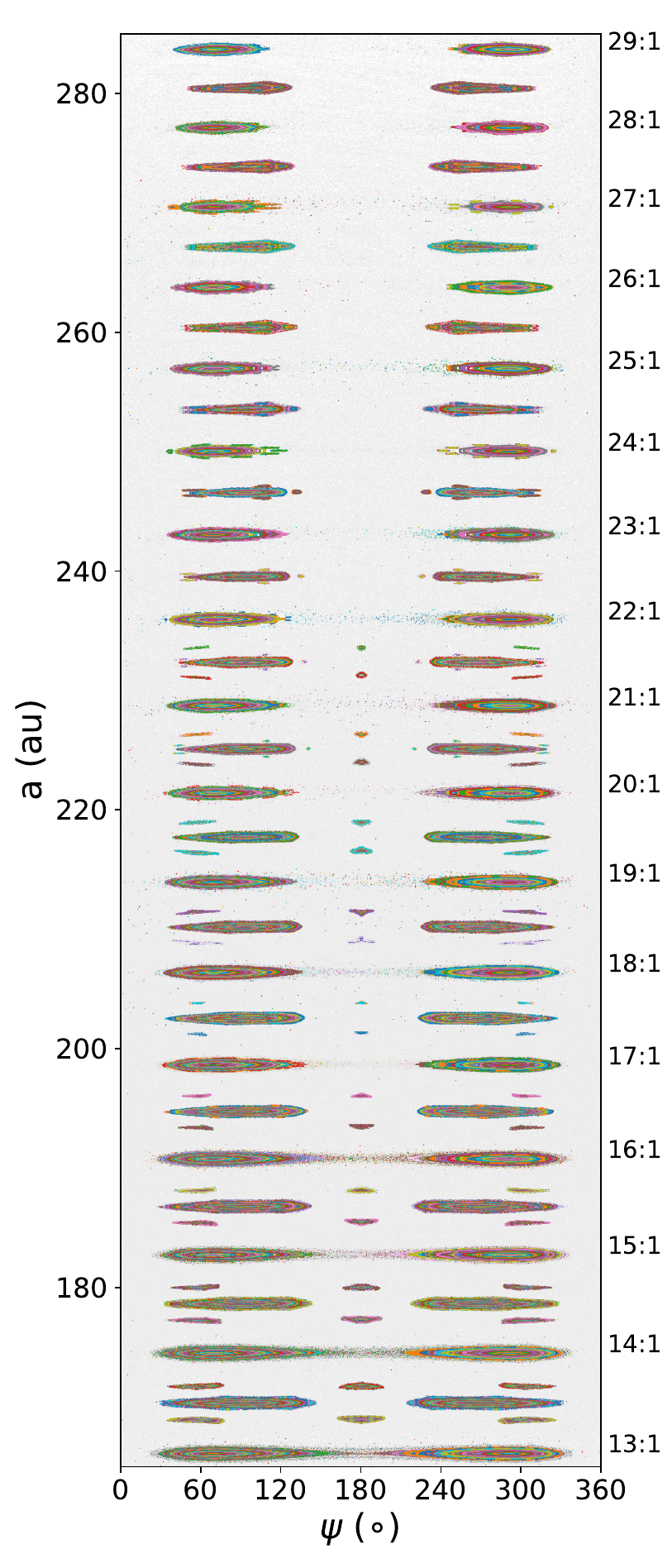} & \includegraphics[width=2.75in]{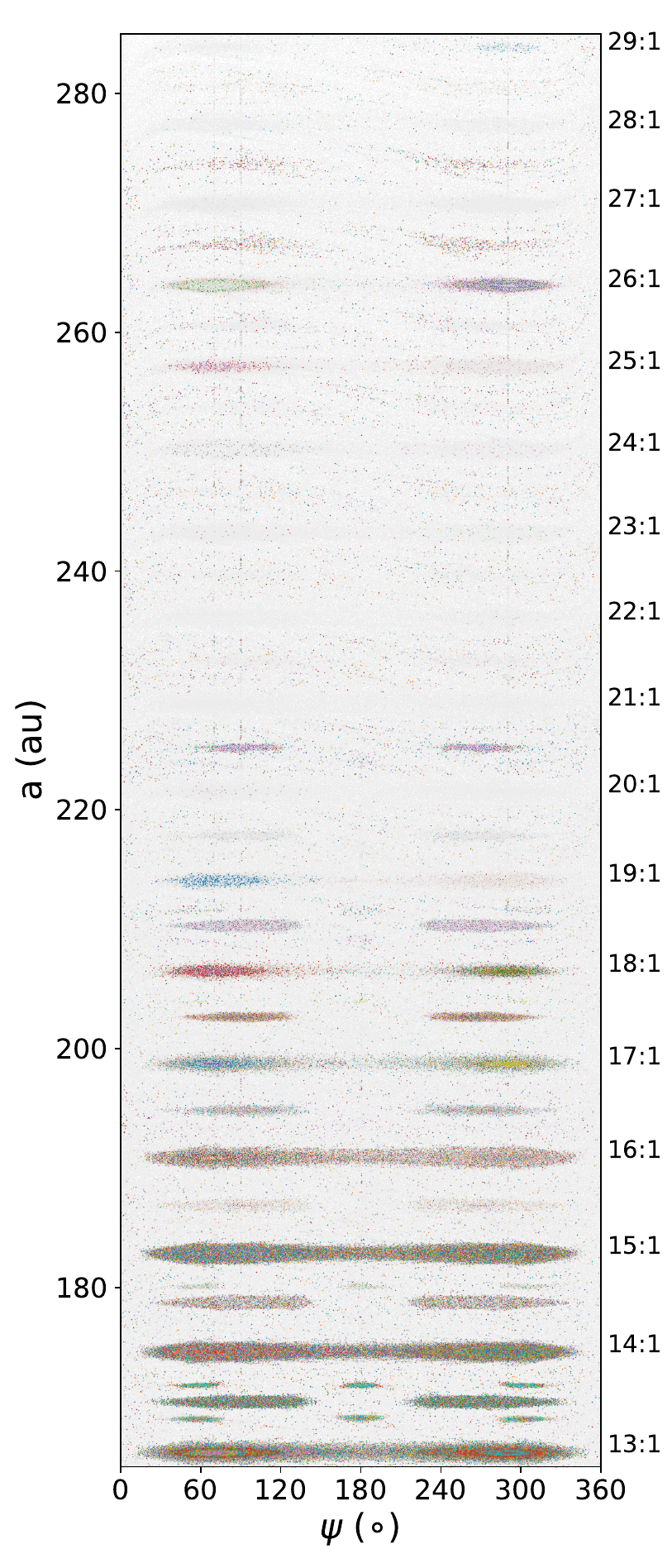} \\
    \end{tabular}
    
    \caption{{Barycentric} semimajor axis (a) vs. $\psi$ for particles in Neptune’s distant resonances, with perihelia of 33 au; {each resonant particle is plotted in a different color while the background scattering population is plotted in gray.} 
    On the left, only the Sun and Neptune are included as perturbers.
    On the right, the Sun and all four giant planets are included as perturbers.
    Note how the N:1 resonances in the 4-planet problem get weaker as the semimajor axis increases and begin to drop out entirely at the 20:1 position, with a brief resurgence around the 26:1 position. This contrasts the more gradual shrinking of distant resonances in the three-body problem}
    \label{fig:3-body-vs-4-planet-mosaic}
\end{figure*}

We start by comparing the $a$-$\psi$ evolution of particles in Neptune's resonances in the simple Sun-Neptune-TNO problem to those in simulations with all four giant planets.
To do this, we created an illustrative $a$-$\psi$ mosaic from 165 au to 285 au in the circular restricted three-body problem, as shown in the left panel of Figure~\ref{fig:3-body-vs-4-planet-mosaic}. 
We included the Sun and Neptune as the only perturbers in our simulations, with Neptune on a circular orbit with its current semimajor axis.
We then initialized massless test particles with perihelia of 33 au and inclinations of 0$^{\circ}$ {with respect to Neptune's orbital plane} along lines of constant semimajor axis spanning from $\psi=0^{\circ}$ to $\psi=360^{\circ}$ at the expected location of all the N:1, N:2, and N:3 resonances in this semimajor axis range. 
We ran the simulation forward through 
{approximately 1000 orbital periods for each test particle (the simulation timescale was set to 1000 times the closest N:1 orbital period for each set of test particles across a small semimajor axis range)}
using the \textsc{rebound} \textsc{ias15} integrator \citep{rebound,ias15}, recording the orbit of each particle every time it passed through perihelion. 
The left panel in Figure~\ref{fig:3-body-vs-4-planet-mosaic} shows the position of the particles at each perihelion passage in $a$-$\psi$ space. 
We plotted any non-resonant particles, loosely defined as any that deviated 5 au or more in $a$ from their starting points, in light gray and the particles that remain resonant in darker colors; this highlights the resonant particles against the background chaos of non-resonant particles.

We then generated a similar $a$-$\psi$ map from 165 au to 285 au for Neptune's resonances in the present-day solar system with the Sun and all four giant planets as perturbers for the test particles.
To do this, we queried JPL Horizons for the positions, masses, and velocities of the Sun, Jupiter, Saturn, Uranus, and Neptune.
We then initialized massless test particles with perihelia of 33 au and inclinations of $0^{\circ}$ {(with respect to the ecliptic)} along four lines of constant $\psi$ spanning the entire semimajor axis range: $\psi=70^{\circ}$ and $\psi=290^{\circ}$ (corresponding {approximately to the centers of asymmetric libration for Neptune's N:1 resonances at our chosen perihelion distances}), $\psi=90^{\circ}$ (corresponding to the center of the N:2 resonances), and $\psi=180^{\circ}$ (corresponding to the center of the N:3 resonances).
Initializing particles at constant $\psi$ rather than along lines at the expected resonant semimajor axes is more efficient in the non-idealized problem; we found that with the four choices of $\psi$,  a resolution of 0.06 au in $a$ was sufficient to resolve all the resonance of interest. 
We note that because the planets are not co-planar, test particles initialized with {ecliptic} $i=0$ in the full simulations results in a small inclination dispersion relative to {Neptune's orbital plane. 
The relaxation of the assumption of co-planarity in this full model also necessitates a slight re-definition of $\psi$; in the full model, $\psi$ is the difference in mean longitudes of the test particle and Neptune at the particle's perihelion, $\psi=\lambda - \lambda_N$ (see \citealt{Volk:2022} for details).
We find that the relative sticking strengths of Neptune's resonances for $q=33$~au is insensitive to a reasonably wide range of inclinations, so the specific choice of reference frame for our $i=0$ runs is unimportant (discussed further below).}
As above, we integrated these particles with the \textsc{ias15} integrator and recorded their orbits at every perihelion passage. 
The resulting $a$-$\psi$ map is shown in the right panel of Figure~\ref{fig:3-body-vs-4-planet-mosaic} with the same color scheme as above.

This comparison reveals that the overall structure of Neptune's resonances is significantly affected by the presence of the other giant planets.
Most notably, the strengths of the N:1 resonances drop off sharply by $a\sim220$~au, which is where the 20:1 would be located, before briefly reappearing at $a\sim260$~au.
Hand-measured widths of the stable libration zones of the N:1 resonances in the full simulations show that they are significantly weaker in this region when compared to the same measurements from the simplified simulations (as seen when comparing the left and right panels of Figure~\ref{fig:3-body-vs-4-planet-mosaic}).
Additionally, in the full problem the N:2 resonances disappear slightly past the position of the 20:1 at $\sim 225$~au (41:2), and the N:3 resonances drop out at $\sim180$~au (44:3).
In the simplified model, the N:2 resonances continue over our entire semimajor axis range, and the N:3 resonances persist to $a\sim235$~au.

{To ensure that our low-inclination simulation results can be generalized to the wider scattering population, we ran a few additional simulations of Neptune's resonances in the full, four giant planet model with test particles at a range of larger inclinations. 
We find that the resonances are essentially unchanged for inclinations of $10^\circ$. 
For inclinations of $20^\circ$ and $30^\circ$, there is a small amount of enhanced sticking to the most distant resonances in our range; this is consistent with the expected increase in the scattering timescale with Neptune due to the particles not always being near the plane of the planets at perihelion \citep[see, e.g.,][]{DiSisto:2020}. 
The insensitivity of our simulations to test particle inclinations is consistent with modeling done by \cite{Gallardo:2019}, who found that the strengths of exterior N:1 resonances at high eccentricity are particularly insensitive to inclination. 
We thus only explore the low-inclination regime in the rest of the paper.}

\subsection{Identifying what is destabilizing resonances in the distant scattering population}\label{ss:destabilization}

\begin{figure}
    \centering
    \includegraphics[width=3in]{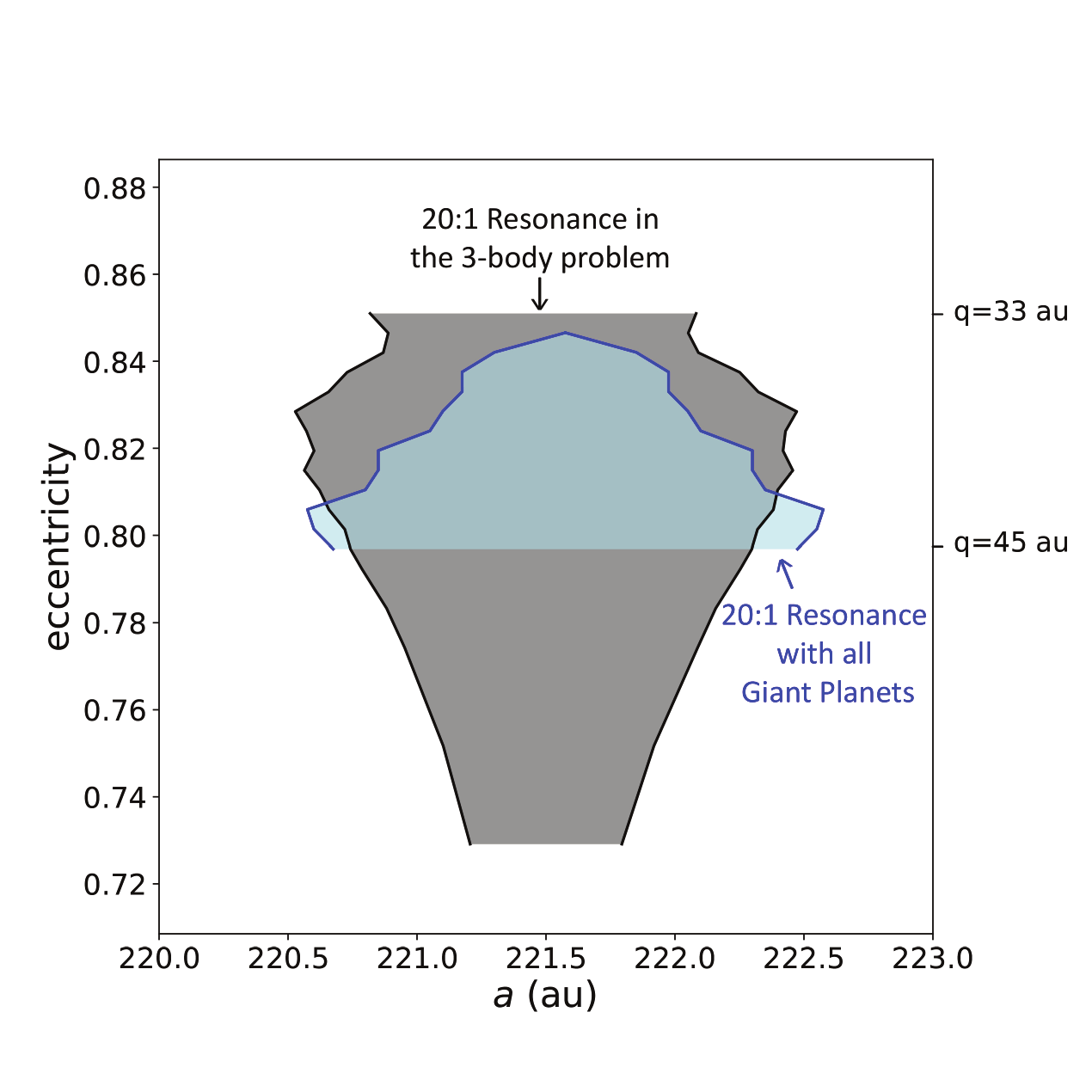}
    \caption{Eccentricity vs. semimajor axis of the 20:1 resonance with Neptune. In gray are resonance widths with only Neptune and the Sun acting as perturbers. In blue are resonance widths with all four giant planets included as perturbers. Notice how resonance width decreases with higher eccentricity in the four-planet case relative to the simplified case.}
    \label{fig:a-e-zoomed}
\end{figure}

To investigate the mechanism responsible for destabilizing Neptune's distant resonances, we consider a case study centered on Neptune's 20:1 resonance, the lowest-$a$ N:1 resonance that lacks a clear stable libration zone in the full simulations (right panel of Figure ~\ref{fig:3-body-vs-4-planet-mosaic}).
We first ran a set of simulations targeting just the 20:1 resonance over a range of perihelion distances $q=33-45$ au in both the simplified model and the full, four giant planet model.
For each value of $q$ (varied in 1 au increments), we initialized 240 massless test particles evenly spaced in $\psi$ from $0^{\circ}$ to $360^{\circ}$ at the center of the 20:1 resonance ($a\approx221.6$~au).
The inclination of all test particles were set to zero; in the simplified model this makes them truly co-planar with Neptune while in the full model they have very small inclinations relative to Neptune.
We again used the \textsc{ias15} integrator to integrate forward over 1000 {test particle orbital periods},
recording the test particles' orbits every perihelion passage to produce $a$-$\psi$ maps. 
The boundaries of the 20:1 resonance's stable libration zones in these $a$-$\psi$ maps of both the simplified and full models were hand measured and are plotted in $a$-$e$ space in Figure~\ref{fig:a-e-zoomed} to show the width of the 20:1 resonance as a function of $e$; this figure also shows an extended eccentricity range to higher-$q$ for the simplified model (which is much less computationally intensive to run).
While the maximum width of the 20:1 resonance is similar in both models, its location {in $q$} is shifted to higher {values} in the full model compared to the Neptune and Sun only case {(the resonance is widest at $q=38$~au when only the sun and Neptune are included in the simulation whereas it is widest at $q=44$~au in the full model)}.
There is a significant decrease in the width and strength of the 20:1 resonance at low $q$ when all giant planets are included in the simulation.

This led us to investigate whether one of the three giant planets other than Neptune is destabilizing the 20:1 resonance at higher eccentricities.
Using the same simulation set up as described above in Section~\ref{ss:exploration} for generating the four-planet model $a$-$\psi$ maps, we ran six simulations of Neptune's resonance in the semimajor axis region surrounding the 20:1 with the following objects included:
\begin{itemize}
    \item Sun + Saturn + Uranus + Neptune + TNOs
    \item Sun + Jupiter + Uranus + Neptune + TNOs
    \item Sun + Jupiter + Saturn + Neptune + TNOs
    \item Sun + Jupiter + Neptune + TNOs
    \item Sun + Saturn + Neptune + TNOs
    \item Sun + Uranus + Neptune + TNOs
\end{itemize}
where in each case the Sun and planets are massive perturbers and the TNOs are massless test particles.

\begin{figure*}
    \centering
    \includegraphics[width=3in]{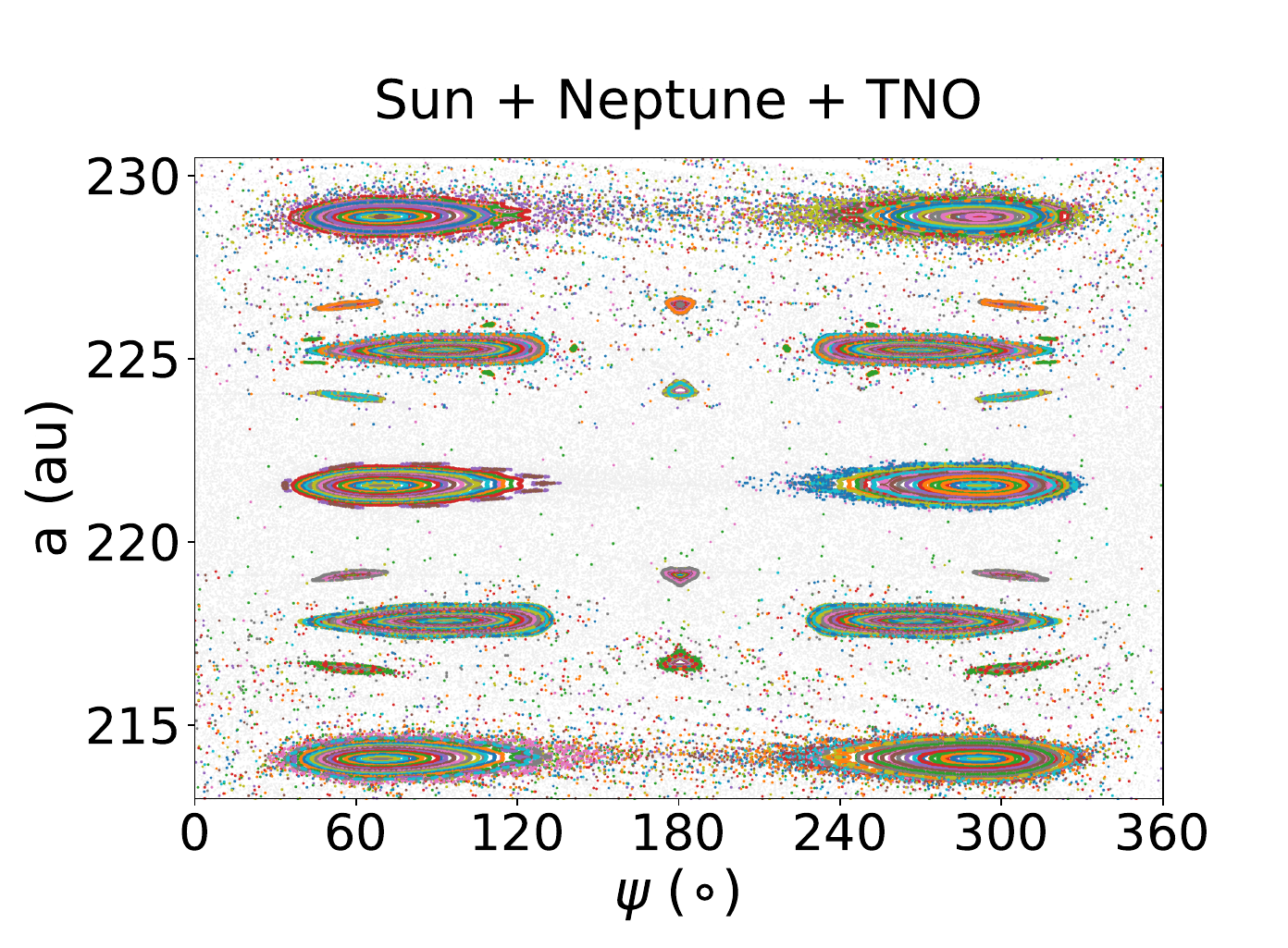} 
    \begin{tabular}{c c}
    \includegraphics[width=3in]{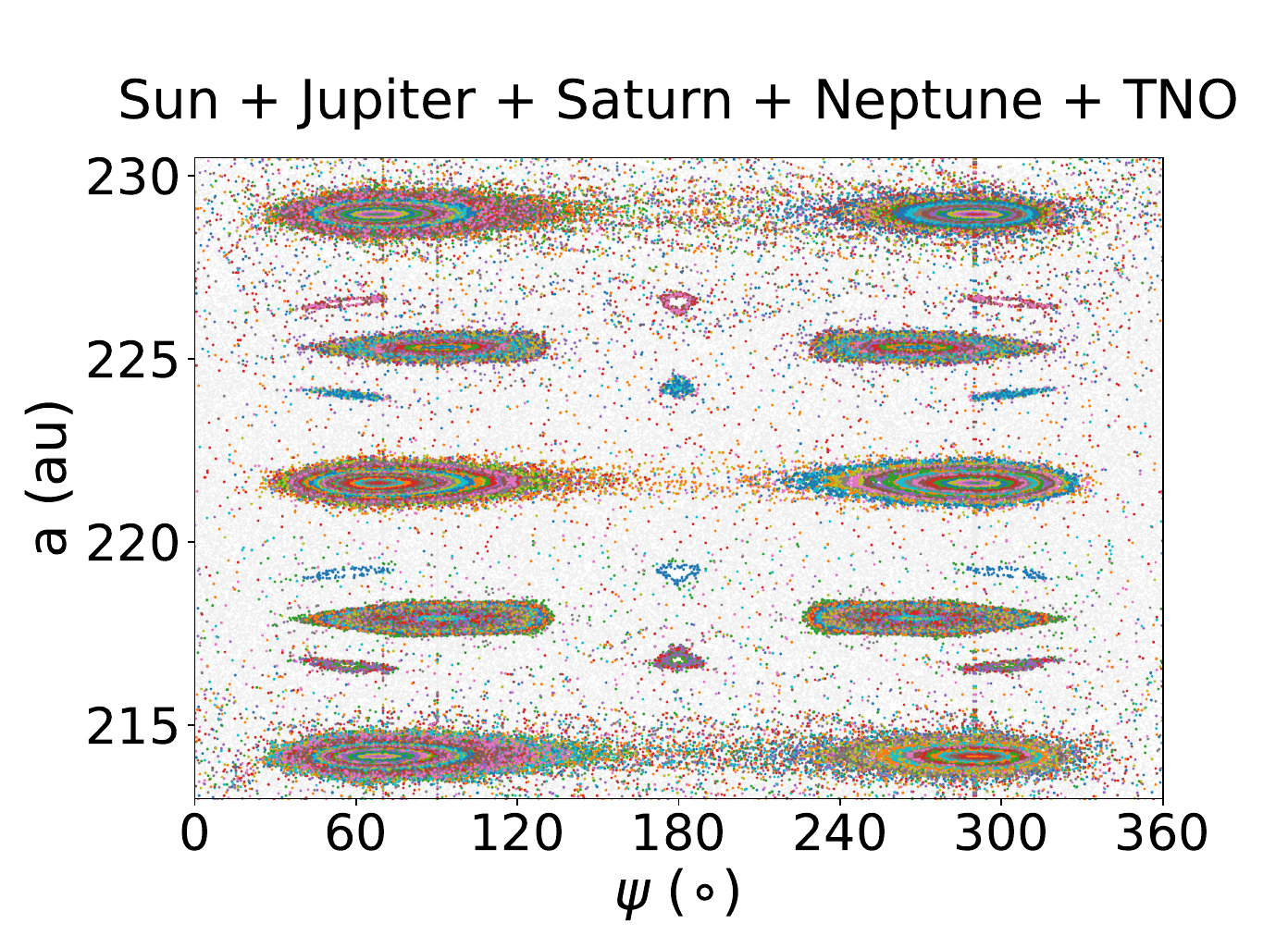}&
    \includegraphics[width=3in]{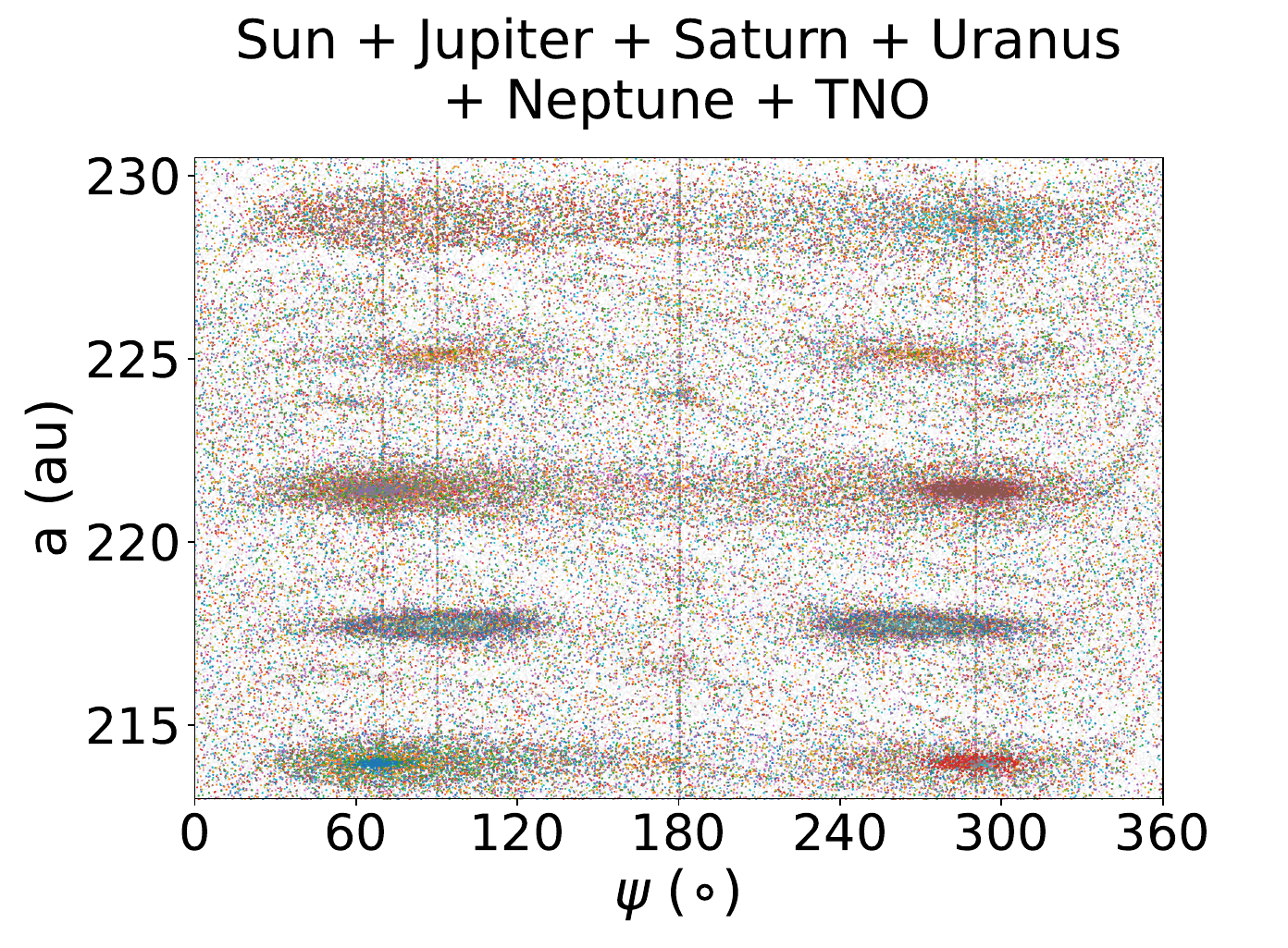} \\
    \includegraphics[width=3in]{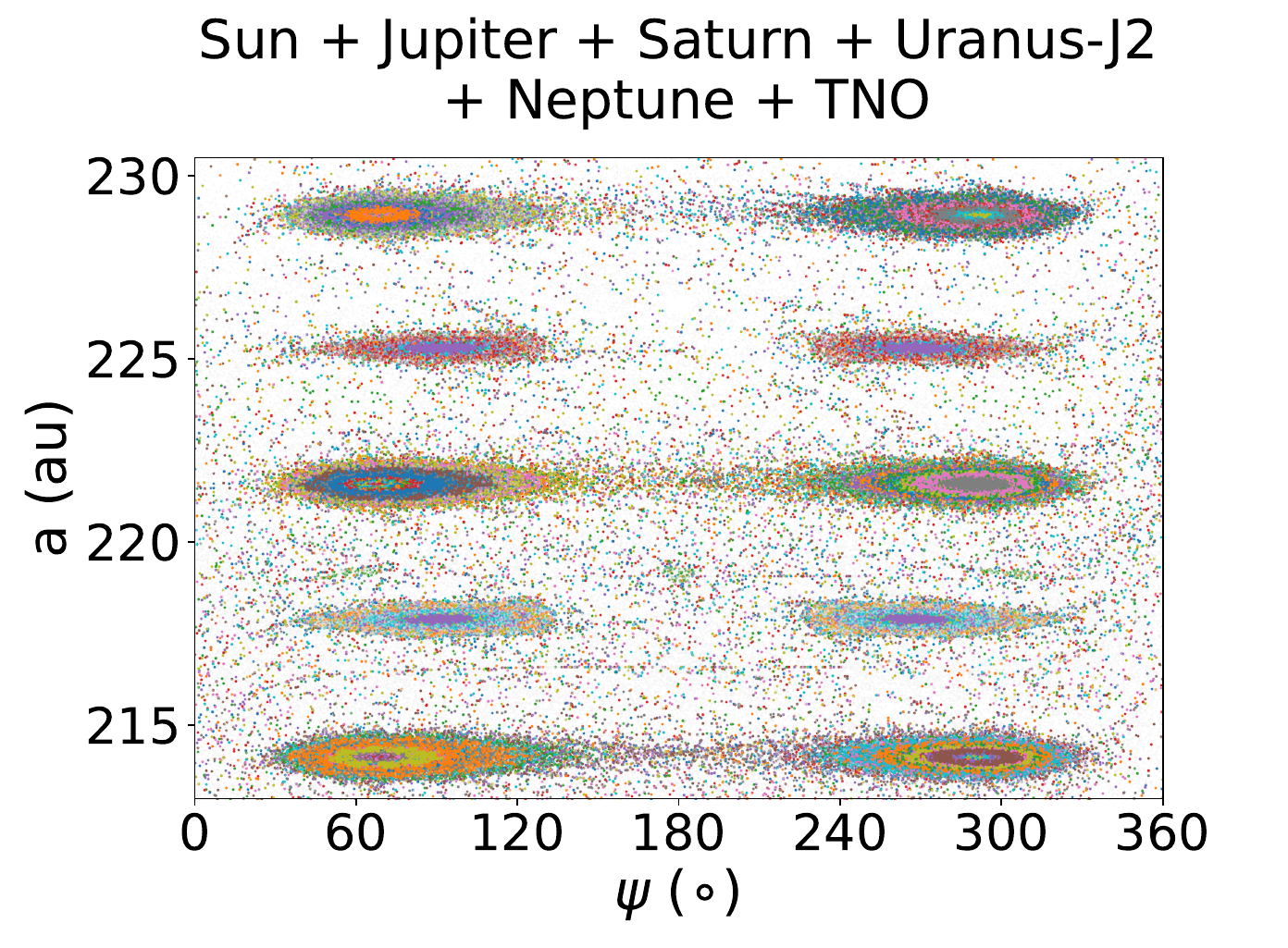} &
    \includegraphics[width=3in]{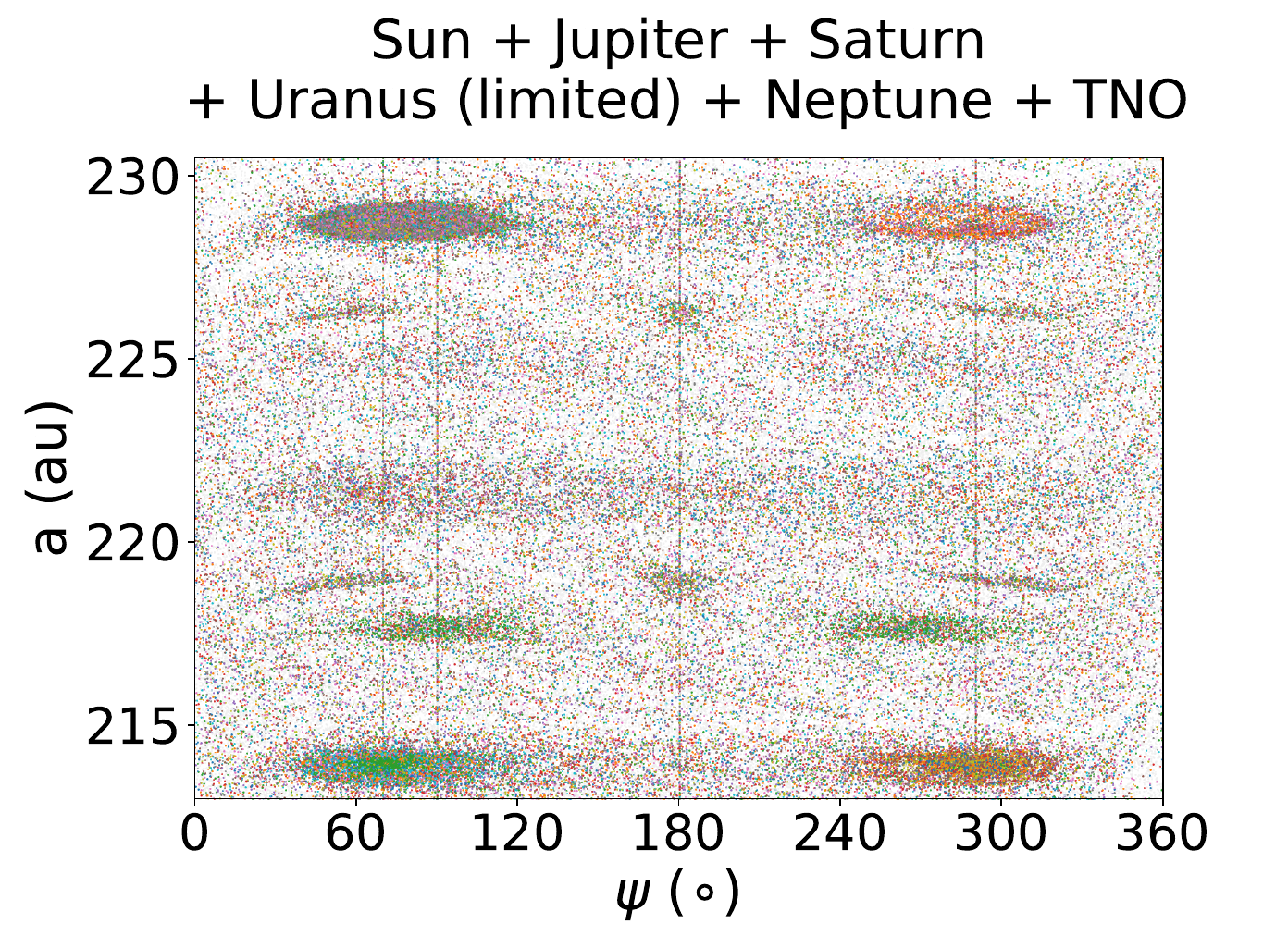}
    \end{tabular}
    \caption{Barycentric semimajor axis ($a$) vs. $\psi$ for particles near Neptune’s 20:1 resonance ($a\approx221$~au) with perihelion distance $q=33$~au. 
    In all panels, we plot particles that remain within 10~au of their initial semimajor axis in colored points, with each particle plotted in a different color (these are the particles that tend to interact with the resonances); less stable particles that quickly scatter away from their starting $a$ are plotted as gray background points.
    In the top panel, only the Sun and Neptune act as perturbers on the test particles.
    In the middle left panel, the Sun and all major planets except for Uranus act as perturbers.
    In the middle right panel, the Sun and all 4 major planets and the Sun act as perturbers.
    In the bottom left panel, all giant planets except Uranus act as perturbers, with the addition of a J2 parameter to simulate Uranus’ secular effects.
    In the bottom right panel, the Sun and all four giant planets are included as perturbers in the integration of the massive bodies, but the test particles are only perturbed by the Sun, Jupiter, Saturn, and Neptune.
    Note the similarity between the simplified (Sun + Neptune) model at the top and the middle left panel with Uranus excluded from the simulation. 
    The destabilization of the 20:1 resonance only occurs when Uranus is included in the simulation; the destabilization remains when Uranus perturbs Neptune's orbit but is not included as a perturber for the test particles themselves.}
    \label{fig:20-1-evolution}
\end{figure*}

In doing this, we were able to isolate the effects of each {planet's short-term perturbations} on Neptune's resonances.
We found that when just Jupiter and/or Saturn were removed from the simulation, the 20:1 resonance remained weak at $q = 33$ au.
However, whenever Uranus was removed from the simulation, the 20:1 resonance was nearly as strong as predicted by the simple three-body problem and did not drop out relative to the neighboring resonances.
We also ran a simulation with the Sun, Jupiter, Saturn, and Neptune as massive perturbers but added a J2 term to represent the averaged secular effect of Uranus on the other planets and the TNO test particles (we used \citealt{Malhotra:2022}'s J2 value and formulation for the simulation modifications).
The 20:1 resonance in this simulation behaved very similarly to the simulation without Uranus, though some of the higher-order resonances nearby appear less distinct. 
Figure~\ref{fig:20-1-evolution} shows a comparison between a few of these test simulations.

From these tests, we can conclude that the weakening of the 20:1 resonance at $q = 33$ au is related to the presence of Uranus, as the resonance is weak whenever Uranus is present in the simulation, but unaffected by the presence of Jupiter or Saturn.
We also conclude that the weakening is not purely or dominantly due to the secular effects of Uranus. 
It is clear that gravitational perturbations by Uranus can have a significant impact on Neptune’s distant resonances for high eccentricity TNOs--a regime typically assumed to be dominated only by the gravitational effects of Neptune.
We ran one more test simulation near the 20:1 resonance to examine whether the weakening of the resonances when Uranus is present is due to Uranus's direct influence on the test particles (via, e.g., resonant interactions with Uranus) or is due to Uranus's influence on the other giant planets' orbits; as we will discuss in Section~\ref{sec:NUperiod}, Neptune and Uranus are near a mutual resonance, so Neptune's orbital evolution is influenced significantly by non-secular effects from Uranus.
In this test simulation, we modified \textsc{rebound}'s \textsc{whfast} integrator \citep{whfast} such that Uranus was included as a perturber for the Sun and the other giant planets, but was not included in the force calculations for the test particles\footnote{We chose to use \textsc{whfast} for this modification because it involved just a very simple alteration to the `kick' subroutine for the test particles within the drift-kick-drift scheme for the \textsc{whfast} algorithm. We ran a test to ensure that the \textsc{whfast} results for a full Sun plus four giant planets plus test particles simulation are comparable to the results of our \textsc{ias15} simulations. The switch in integrator for this test does not affect the results.}.
The results of this test simulation are also shown in Figure~\ref{fig:20-1-evolution}, and the resonance structures are the most similar to the full Sun plus all four giant planets simulations. 
This {suggests} that, near the 20:1 resonance at least, the weakening of Neptune's resonances is caused by variations in Neptune's orbit due to its near resonance with Uranus. 

We explore in more detail how Uranus is affecting a wider range of Neptune's resonances in Section~\ref{sec:NUperiod}.  
But, first, we must establish a more quantitative method to measure resonance strengths in our simulations.

\subsection{Quantifying resonance strengths}\label{ss:strengths}

Previously, we have used resonance width (in semimajor axis) to characterize the strength and stability of Neptune’s external resonances. 
This is a useful metric in many circumstances, such as in the Sun-Neptune-TNO three-body problem. 
In the three-body case, Neptune's resonances {generally have} clear boundaries in $a$-$\psi$ phase space. 
These well-defined boundaries break down and become more chaotic when all giant planets are included in our simulations. 
This makes identifying the boundaries of resonances in the four-planet model more subjective, as the dimensions change depending on the parameters used to plot the test particles (such as limiting test particles by change in semimajor axis over a specified simulation time). 
This is especially difficult when studying scattering (low-$q$) populations. 
Because of this we sought a different, more concretely defined, metric for measuring the strength and stability of resonances in the full six-body problem (Sun + four giant planets + TNO).

\begin{figure*}
    \centering
    \includegraphics[width=5in]{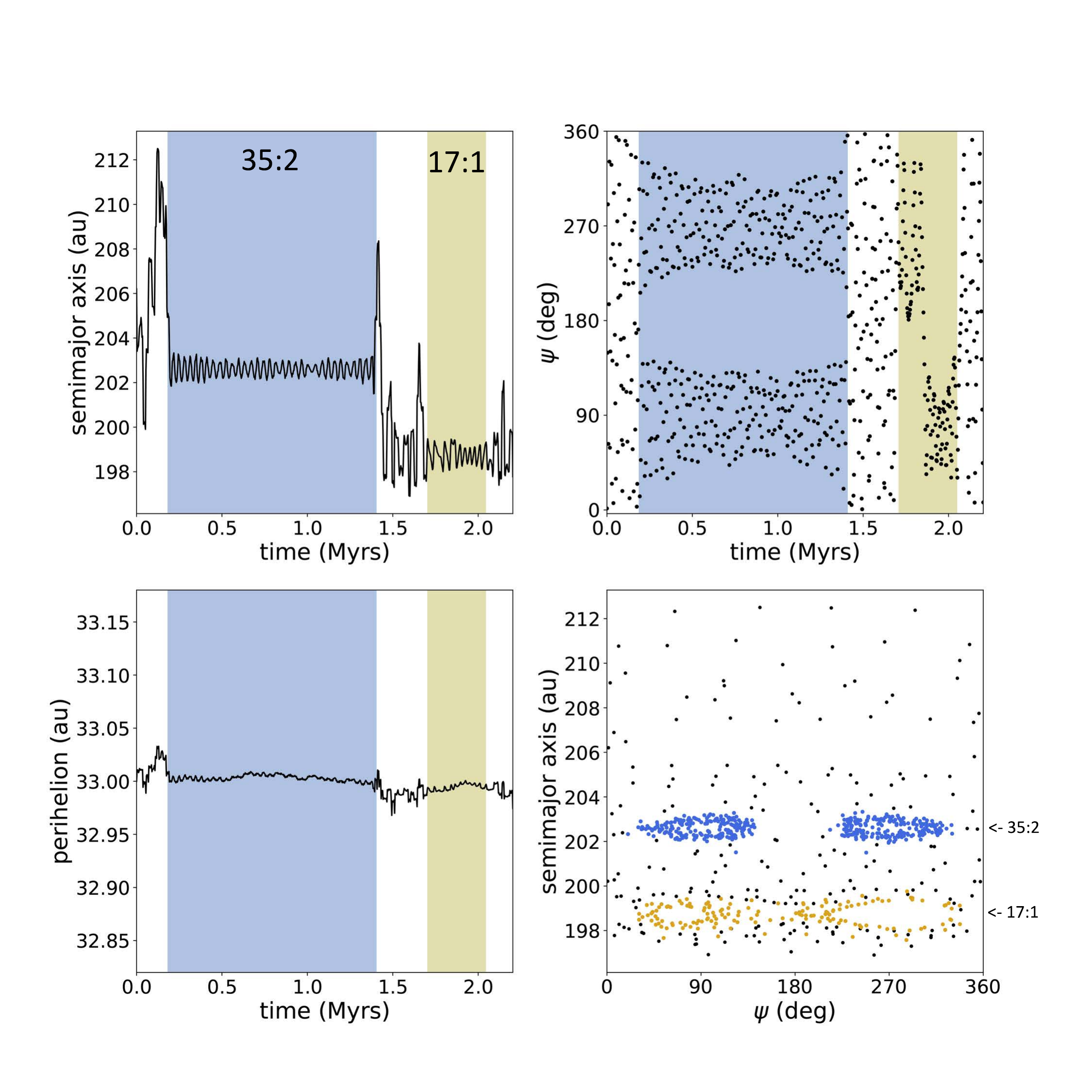}
    \caption{Example evolution of a test particle that scatters and sticks to Neptune’s resonances.
    We show the evolution in {barycentric} $a$ vs time (upper left), $\psi$ vs time (upper right), $q$ vs time (lower left) and in $a$-$\psi$ space (lower right).
    We highlight sticking events in the 35:2 ( $\psi$ alternates between $\sim90^\circ$ and $\sim270^\circ$ {at each perihelion passage}) and the 17:1 ($\psi$ first librates in the leading asymmetric island then the trailing island).}
    \label{fig:sticking-particle-ex}
\end{figure*}

The following describes the basic behavior of some typical low-$q$ test particles simulated in the full six-body case.
Test particles initialized near the edge of more stable, closer-in resonances will often move in and out of resonance over the span of a simulation due to chaotic gravitation interactions with the giant planets.
 Therefore, these test particles cannot be defined as explicitly in or out of resonance, but rather are transient.
This is one difficulty when it comes to measuring widths of resonances in the six-body problem: do we count transient particles as resonant or not?
In the case of further out resonances (such as the 20:1 at $q = 33$~au and beyond), there may be no simulated test particles that are persistently resonant  over the timescale we are considering. 
But that does not mean there is no resonant interaction, and it would be inaccurate to say that these resonances have a width or strength of zero.
So somehow, we must account for test particles that transiently pass in and out of, or stick, to each resonance; example typical sticking behavior is demonstrated in Figure ~\ref{fig:sticking-particle-ex}.
Additionally, even test particles that spend most of their time librating in a resonance will often `slip' in $\psi$ at chaotic boundaries, so purely considering confinement of $\psi$ doesn't provide a full measure of the stability and strength of a resonance
Overall, using the width to measure these resonances based only on long-term stable libration does not accurately capture the strength of these sticking interactions that are very important in the scattering TNO population.

In these situations, it is far more useful to look at the total resonant lifespan of test particles in a given simulation {(see, e.g., \citealt{Lykawka:2007})}.
That is, how long do test particles spend interacting with a given resonance as a fraction of the total simulation time, even if these interactions are not continuous.
We quantify this fractional resonance occupation time averaged over all the test particles in a given region around a resonance as our resonance strength parameter (exact details of its calculation are provided in Appendix~\ref{sec:appendix}). Figure~\ref{fig:strengths} shows the relative strengths of Neptune's distant resonances at $q = 33$~au using this metric. This method is also utilized in the next section to quantify resonance strength.

\begin{figure}
    \centering
    \includegraphics[width=3in]{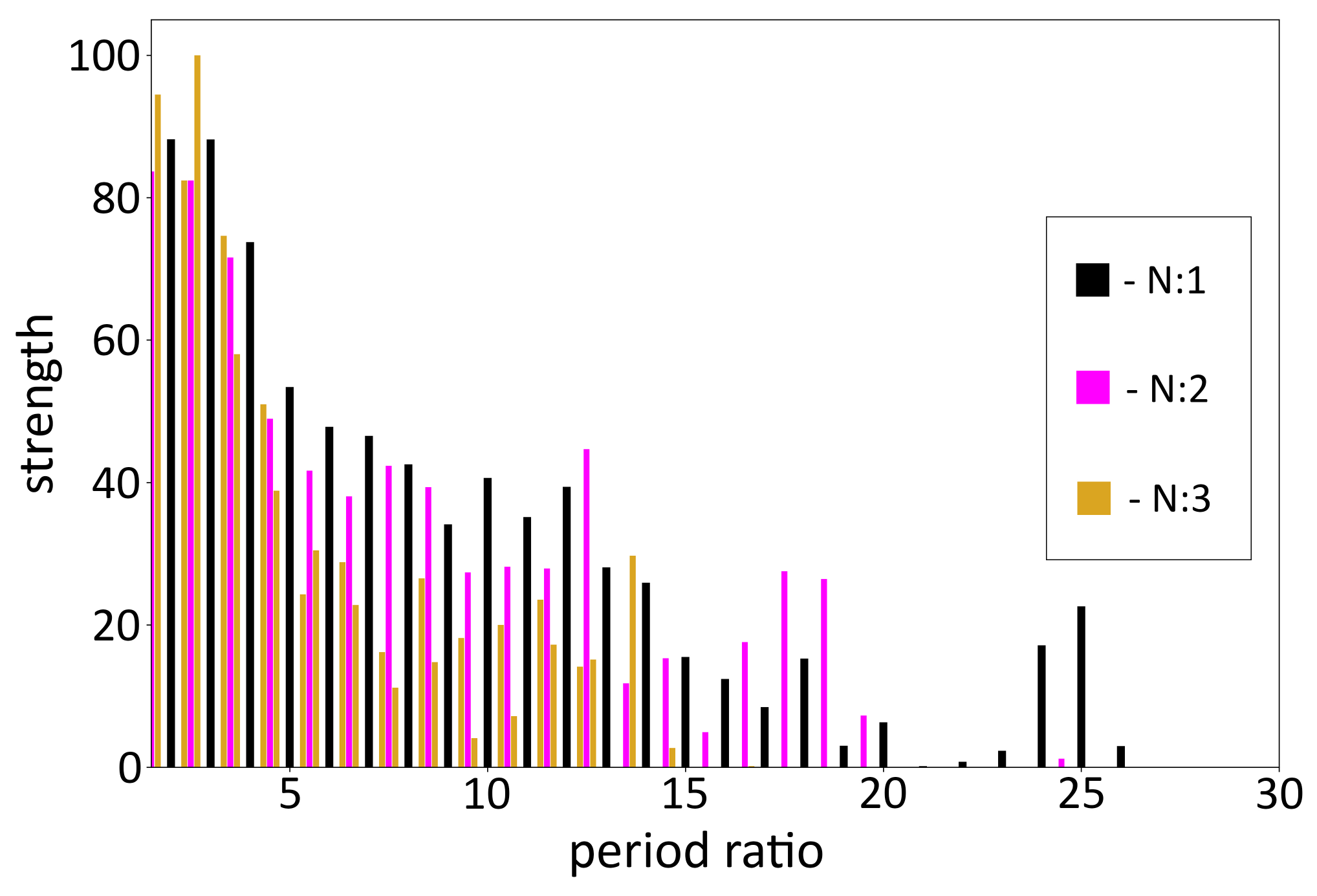}
    \caption{Relative strength of the N:1 (black), N:2 (pink), and N:3 (yellow) resonances with Neptune for low-inclination particles with $q = 33$~au in a model of the current solar system (with the Sun and all four giant planets as perturbers). Our strength metric is normalized to the largest value across all resonances and is a measure of how confined test particles are to each resonance (see Appendix~\ref{sec:appendix}).}
    \label{fig:strengths}
\end{figure}

The dramatic drop-off in our resonance strength metric is consistent with scattering population simulations that show a dramatic reduction in resonance sticking at large semimajor axes ($a\gtrsim250$~au; \citealt{Lykawka:2007}).
We have established that Uranus is directly linked to this drop-off in resonance strength (Figure~\ref{fig:20-1-evolution}).
Neptune's current orbital period is 1.96 times Uranus's orbital period, which is close to their mutual 2:1 resonance.
In the next section, we explore how this near-resonance might contribute to weakening Neptune's distant external resonances.

\section{Experimenting with the Neptune-Uranus period ratio} \label{sec:NUperiod}

As discussed in Section~\ref{ss:destabilization}, it is clear that Uranus has a significant influence on Neptune's distant resonances, and there are hints that a lot of that influence is due to how Uranus affect's Neptune's orbital evolution. 
Uranus and Neptune are not far from a mutual 2:1 resonance, so here 
we we explore how the proximity to that mutual resonance can reduce stability in Neptune's distant resonances. 
To test this we ran a series of simulations with Neptune’s semimajor axis shifted slightly from its present day value to change its mutual orbital period ratio with Uranus.
We chose to simulate the phase space between Neptune’s 2:1 and 28:1 resonances at {the equivalent of} $q = 33$~au, for Neptune to Uranus (N:U) period ratios {from 1.76 out to 2.36, far from their mutual 2:1.}

For each N:U period ratio simulation we included all four giant planets with their orbital parameters being queried from JPL Horizons in barycentric coordinates.
We then changed only Neptune’s semimajor axis to correspond with our desired N:U period ratio.
All of Neptune’s other orbital parameters were left at their present day values, and we recomputed the new barycenter of the simulated system.
Before adding test particles to this modified system, we made sure that the giant planets' orbits remained stable, {particularly Neptune's semimajor axis and eccentricity, for the 4.6~Myr maximum duration of our resonance simulations.}

For each N:U period ratio, we then ran a set of simulations with test particles spanning the expected locations of Neptune's N:1 resonances from the 2:1 to the 28:1, with a $\pm0.5$ period ratio range around each N:1 to include the nearby higher-order resonances.
For the N:U=1.96 simulation (Neptune’s current day position), we assigned all particles a perihelion distance of 33 au to match our previous simulations of scattering bodies.
When moving Neptune’s semimajor axis to change the N:U period ratio, we scaled the test particle perihelia; {e.g, 
$q=33.3$~au with Neptune at $a = 30.37$~au (N:U$= 2.00$),  $q = 33.65$~au with Neptune at $a = 30.65$~au (N:U$= 2.02$), and $q = 37.3$ au with Neptune at $a = 34$~au (N:U$= 2.36$).}
For each simulation, the 1000 particles were allocated uniformly across the N$\pm0.5$ period ratio range along lines of constant $\psi$ values as follows: 500 particles targeted the centers of the asymmetric N:1 resonant islands (250 at $\psi=70^\circ$ for the leading island, and 250 at $\psi=70^\circ$ for the trailing island), 250 targeted the centers of the N:2 resonances at $\psi=90^\circ$, and 250 targeted the centers of N:3 resonances at $\psi=180^\circ$.
The particles were initialized at perihelion (mean anomaly of zero), so their longitude of perihelia were determined by the desired $\psi$ value.
Their semimajor axis and eccentricity were set by the period ratio with Neptune and fixed perihelion distance.
We set inclinations and longitudes of ascending node to zero; as noted in Section~\ref{ss:exploration}, because the planets are not co-planar in our simulations, this amounts to allowing the test particles to have a small inclination dispersion relative to Neptune (a few degrees at most over the timescale of our simulations).
We found that these initial parameters for test particles were adequate to resolve the general structure of Neptune’s resonances when plotted in phase space and to calculate resonance strengths.

\begin{figure*}
    \centering
    \begin{tabular}{c c c c c}
        \includegraphics[width=1.375in]{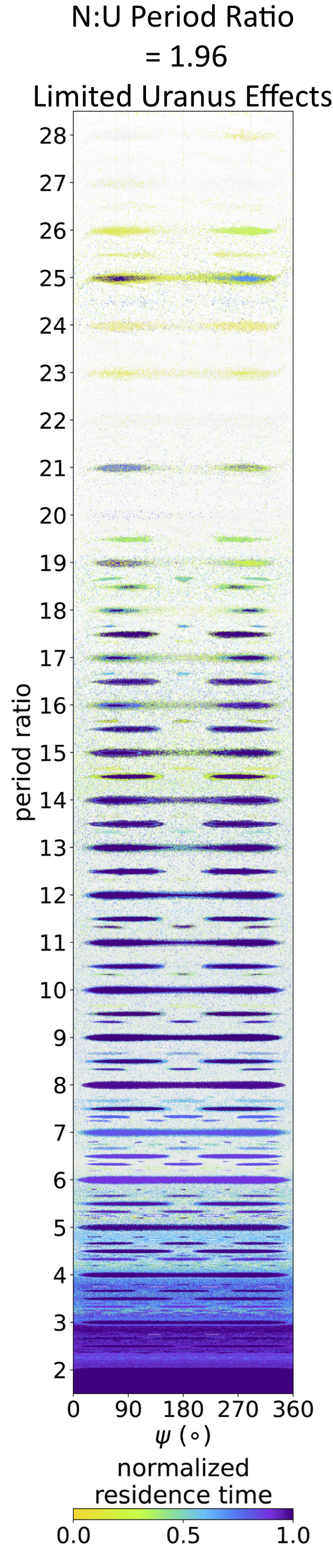} &
        \hspace{-10pt}\includegraphics[width=1.375in]{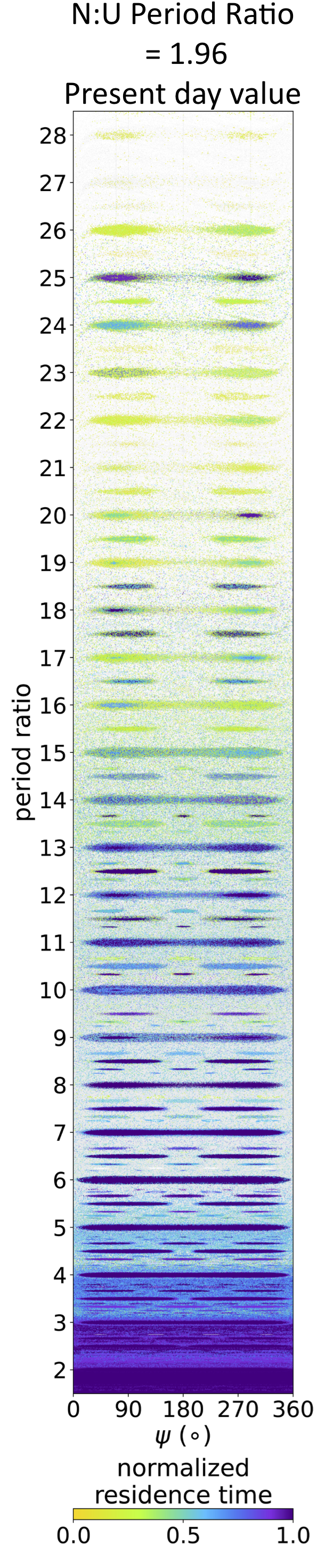} & 
        \hspace{-10pt}\includegraphics[width=1.3735in]{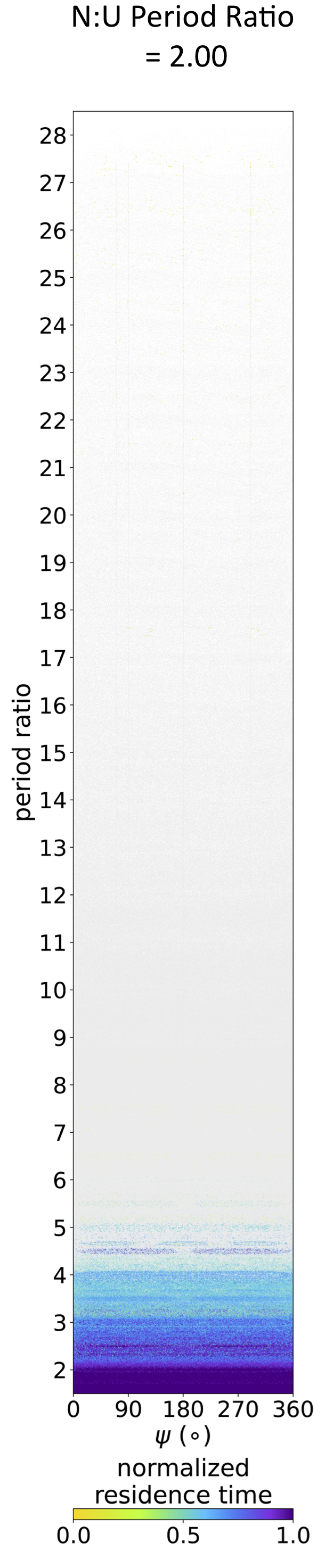} &
        \hspace{-10pt}\includegraphics[width=1.375in]{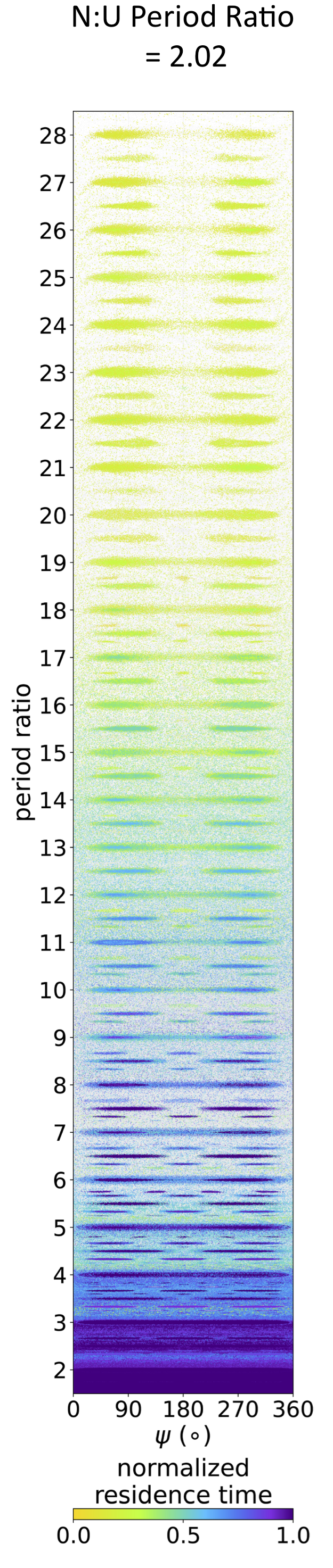} &
        \hspace{-10pt}\includegraphics[width=1.375in]{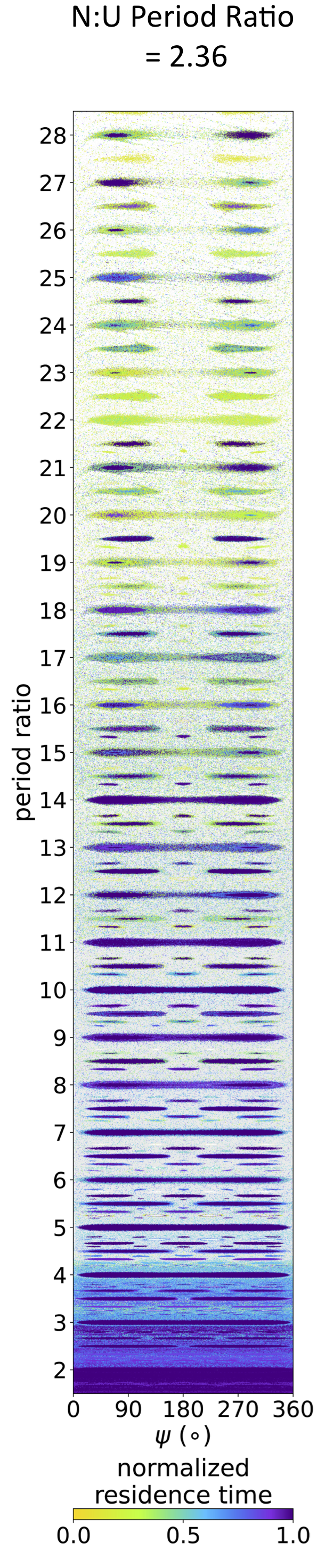} \\

    \end{tabular}
    
    \caption{Test particle period ratio with Neptune vs. $\psi$ in the phase space from Neptune’s 2:1 to 28:1 resonances ($\sim50-280$~au) for four different Neptune : Uranus period ratios. In each simulation, particles are initialized with the equivalent of $q=33$~au for present-day Neptune's orbit. Dark blue test particles indicate a longer normalized residence time, which is defined as a function of its change in semimajor axis ($\Delta$a) across multiple small time intervals (see Appendix~\ref{sec:appendix}). 
    Lighter yellow test particles indicate a shorter normalized residence time. Particles assigned a residence time of 0 (i.e., particles that scatter for the entire simulation) are plotted in light gray for clarity to separate them from the resonant-interacting particles. 
    Uranus is included as a full perturber in all but the left-most plot; in the limited Uranus effects simulation, Uranus interacts with the other massive bodies but not directly with the test particles.  
    Notice the dramatic destabilization of Neptune's external resonances when Neptune is set to be in the 2:1 resonance with Uranus. We discuss in Section~\ref{sec:NUperiod} how the stability of Neptune's external resonances depends on the the mutual period ratio between Uranus and Neptune.
    In the leftmost figure, note that the destabilization remains when Uranus perturbs Neptune, but is not included in the force calculations for the test particles.}
    \label{fig:4-planet-mosaic}
\end{figure*}

\begin{figure}
    \centering
    \includegraphics[width=3.5in]{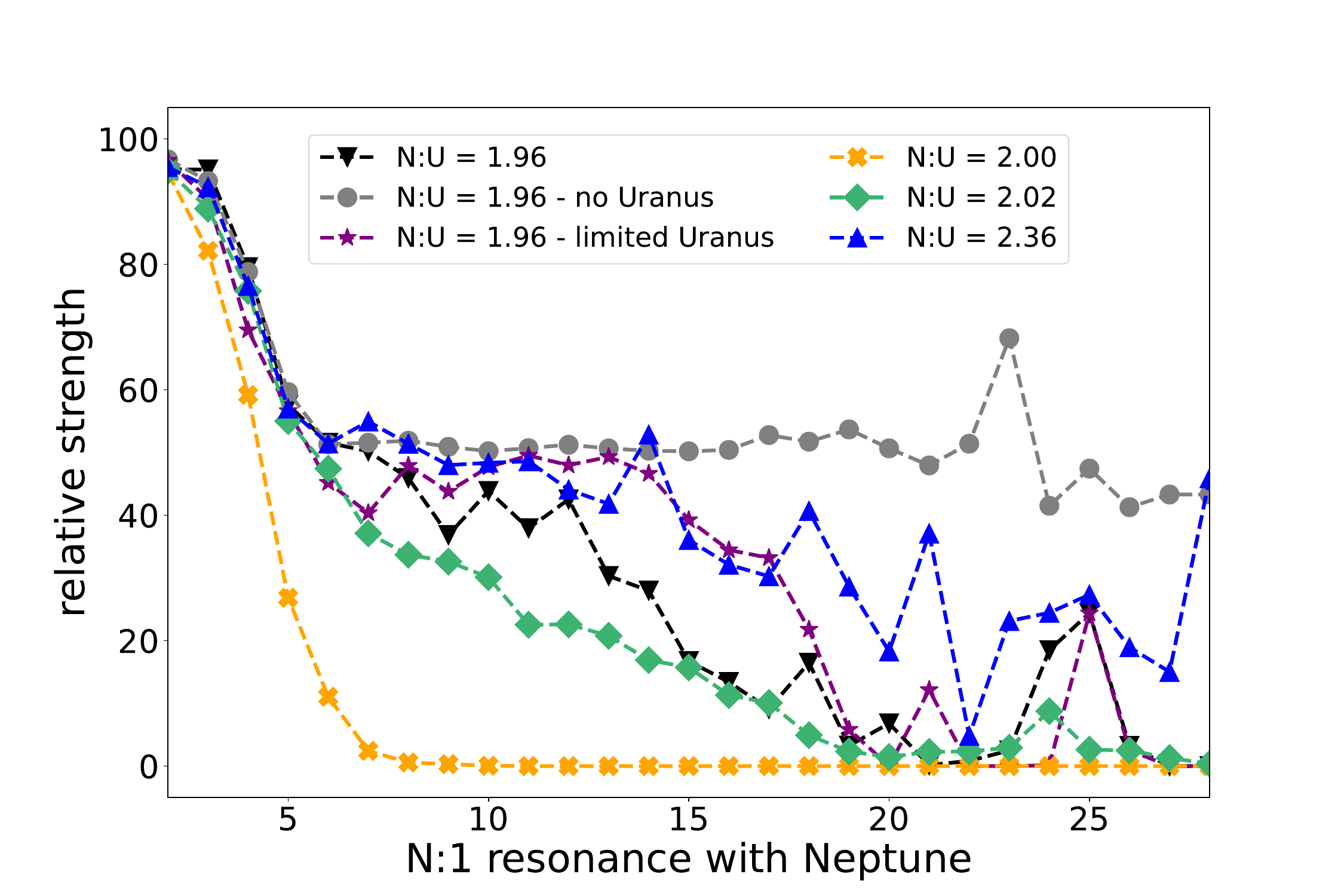}
    \caption{Relative sticking strengths of Neptune's N:1 resonances for different Neptune : Uranus period ratios {see Appendix~\ref{sec:appendix} for the definition of our strength parameter)}.
    The black triangles and dashed line show the strengths in the current solar system (Sun plus all four giant planets on their present-day orbits); the gray circles and dashed line shows the strengths with Uranus removed. 
    The four sets of colored points and lines show strengths from simulations with Jupiter, Saturn, and Uranus on their current orbits, and Neptune's semimajor axis adjusted to the labeled period ratio with Uranus (these are the same simulations shown in Figure~\ref{fig:4-planet-mosaic}).
    The purple colored points and lines show strengths from simulations where Uranus is included as a perturber for the other massive bodies but not for the test particles.
    In all simulations, the strengths have been normalized such that Neptune's external 2:1 resonance has a strength of 100.
    There is a very rapid drop-off in N:1 resonance strengths with increasing N at the exact N:U 2:1 resonance.
    The resonance strengths plateau past the $\sim5:1$ in the simulation with Uranus removed.
    In the simulations with Uranus, the destabilization of Neptune's distant N:1 resonances is more pronounced when the N:U period ratio is closer to 2:1. 
    The simulations with limited gravitational effects from Uranus on the test particles show similar destabilization.}
    \label{fig:strength-data-poster}
\end{figure}

As before, we ran these simulations for a duration of 1000  {test particle orbital periods} of the target N:1 resonance in each simulation using \textsc{ias15}. 
We analyzed resonance strengths {for each N:U period ratio} using the methods described in Section~\ref{ss:strengths} and Appendix~\ref{sec:appendix}.
The resulting mosaics of individual particle residence times near the investigated resonances are shown Figure~\ref{fig:4-planet-mosaic} for four {selected} N:U period ratios, and the calculated strengths of each N:1 resonance for the same N:U period ratios are shown in Figure~\ref{fig:strength-data-poster}.

\begin{figure*}
    \centering
    \includegraphics[width=6.5in]{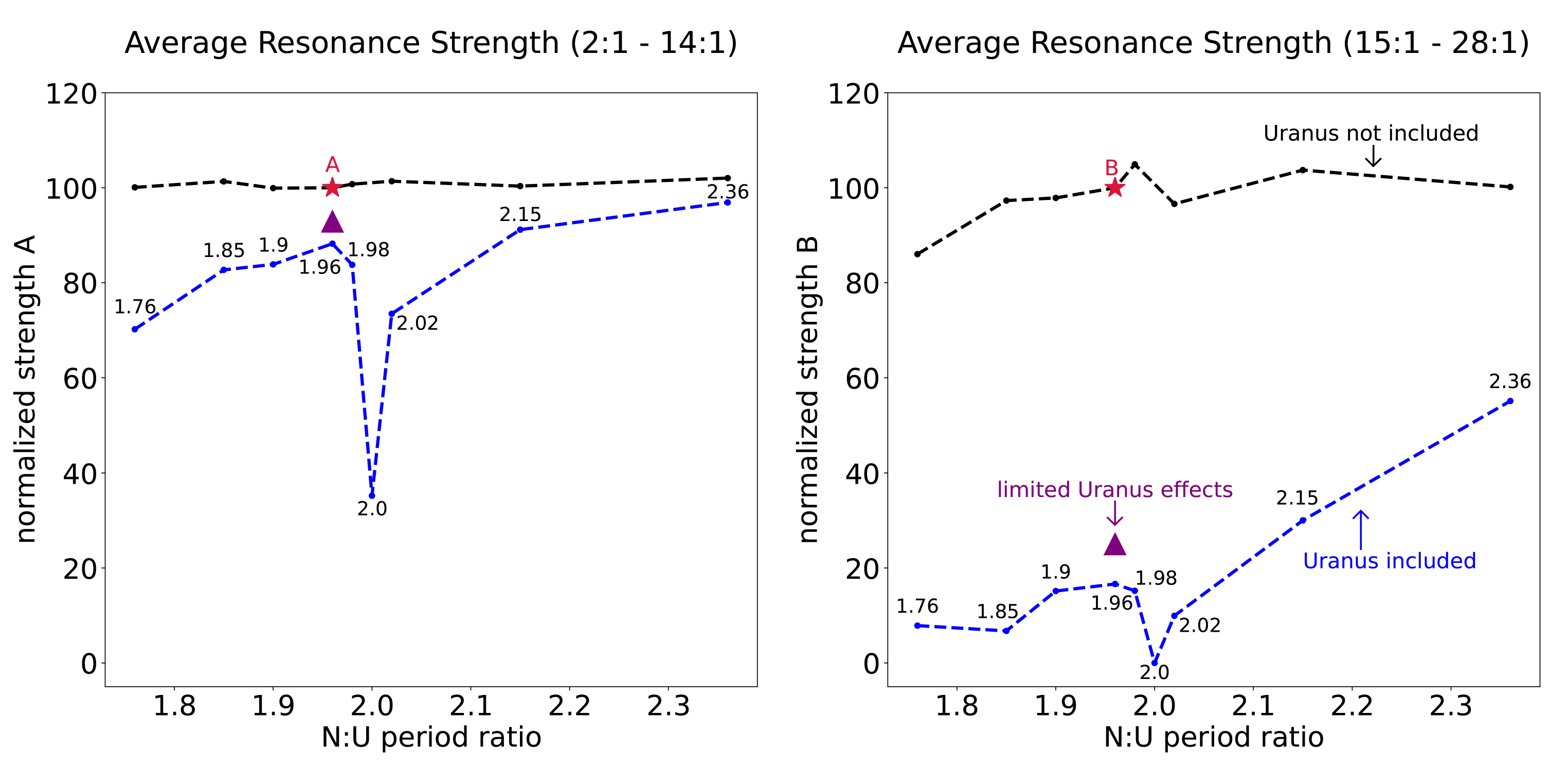}
    \caption{Resonance strengths as a function of the Neptune : Uranus period ratio averaged across Neptune's 2:1 through 14:1 resonances (left panel) and 15:1 through 28:1 resonances (right panel).
    The blue points and dashed lines show the averaged strengths in simulations with the Sun and all four giant planets, with Neptune's semimajor axis increased or decreased to change its period ratio with Uranus.
    The black points and dashed lines show the averaged strengths in simulations with the same values of Neptune's semimajor axes but with Uranus removed (i.e., simulations with the Sun, Jupiter, Saturn, and Neptune as perturbers).
    The red stars highlight the present day N:U period ratio (1.96) in the simulations with Uranus removed; the average resonance strengths measured each simulation are normalized relative to these two values (both red stars are at y=100).
    The purple triangles represent the averaged strength from a set of simulations where Uranus perturbs the orbits of Neptune and the other giant planets, but is not included in the force calculations on the test particles. Note that these values are similar to the those of the 1.96 N:U period ratio simulations with the full effects of Uranus included.
    Neptune's more distant resonances (right panel) are much more dramatically weakened by the presence of Uranus across all tested N:U period ratios.
    All of Neptune's resonances show the most dramatic weakening at period ratios close to the N:U 2:1, and less weakening at higher N:U period ratios. 
    }
    \label{fig:average-strengths}
\end{figure*}

{Examining our full range of simulated N:U period ratios, we find that resonance stability drops off steeply for}
ratios between 1.96 and 2.00 as well as between 2.00 and 2.02.
Between N:U period ratios of 2.02 and 2.36, resonance strength returns steadily. 
To visualize these trends in overall resonance strength with N:U period ratios, we averaged individual resonance strength values across a range of N:1 resonances for each set of N:U period ratio simulations.
{This ``average strength'' value is shown in Figure~\ref{fig:average-strengths} for Neptune's closer-in N:1 resonances (2:1 through 14:1; left panel) and distant N:1 resonances (15:1 through 28:1; right panel)
highlighting the general weakening of all simulated resonances for various N:U period ratios.}  
To provide a baseline for comparison, at each modified value of Neptune's semimajor axis we also ran a set of simulations with Uranus removed from the simulation and calculated the strengths of Neptune's 2:1 through 28:1 in the absence of Uranus's destabilizing effects.

Figure~\ref{fig:average-strengths} shows that the reduction in resonance strength due to Uranus is much more dramatic for the distant resonances than the closer-in resonances, regardless of the specific N:U period ratio.
The weak trend in the baseline simulation's average distant resonance strength (right panel) shows that the other giant planets do influence the strength of Neptune's distant resonances but to a much lesser extent than Uranus.
For both the closer-in and distant resonances, the present day N:U period ratio is actually more favorable for Neptune's resonance strengths than N:U period ratios smaller but farther from their mutual 2:1;
the general proximity of Neptune to Uranus, which increases Uranus's influence on Neptune's orbital evolution, is a significant factor in the destabilization of Neptune's resonances.
It is also clear that if Neptune and Uranus were any closer to their mutual 2:1 resonance than they are today, all of Neptune's external resonances would be significantly weakened. 
We discuss the implications of this further in Section~\ref{sec:discussion}.

We performed a few additional simulations to better isolate which aspect of Uranus's influence on Neptune is the most important, and whether any of the weakening is due to more direct effects of Uranus on the test particles themselves.
One obvious influence Uranus has on Neptune is that it causes Neptune's eccentricity to vary over time, and the amplitude of these variations increases as the N:U period ratio approaches 2:1; we tested whether increasing Neptune's eccentricity variations in the absence of Uranus itself weakened Neptune's resonances.
We ran two sets of simulations with Neptune at its present day semimajor axis and only the Sun, Jupiter, and Saturn as additional perturbers. 
In these simulations, we imposed sinusoidal variations in Neptune's eccentricity and precession of its longitude of perihelion to mimic its present day evolution as well as amplified variations to mimic proximity to the N:U 2:1 without actually including Uranus in the simulation; in the latter case, we based the variation on the N:U$=1.98$ simulations in which Neptune's eccentricity maximum was nearly double its present day value. 
To do this, we used fictitious forces within \textsc{rebound} similar to those described in \cite{Wolff:2012} for the Mercury integrator (the implementation of these integrator modifications within \textsc{rebound} is described in \citealt{Hermosillo-ruiz:2024}).
In both scenarios, we modeled test particles in Neptune's 2:1 through 28:1 resonances, calculated the associated resonance strengths, and visually inspected maps like those shown in Figure~\ref{fig:4-planet-mosaic}.
There were no significant differences in Neptune's resonances as a result of increasing Neptune's eccentricity; both of these test simulations behaved essentially the same as our baseline simulations with Uranus removed.

The other notable aspect of Neptune's evolution that can be attributed to Uranus is that Neptune's barycentric semimajor axis experiences sinusoidal variations with a period of $\sim4300$ years; this is the period associated with the near 2:1 N:U resonance, and the amplitude of these variations are larger when the planets are closer to the resonance.
This period is equal to a bit more than 26 times Neptune's orbital period and thus equal to the orbital periods of TNOs near our most distant set of N:1 resonances where the resonance strengths drop off (though we see the drop-off being closer in than the 26:1).
We note that there is also a longer-period modulation of the amplitude of Neptune's barycentric semimajor axis variations that has a period of $\sim1.15$~Myr, though that is significantly longer than the orbital or resonant libration timescales for the N:1 resonances we consider.
To see whether Neptune's resonances regained any strength at larger distances where the orbital periods are better separated from this short-term period in Neptune's $a$ variations, we extended our present-day solar system configuration simulations out to Neptune's 50:1 resonance; there was no apparent return of resonant stability for particles with $q=33$.

We ran one last set of simulations to determine how much of the observed weakening of Neptune's present-day resonances in the scattering population is attributable just to the Uranus-induced variations in Neptune's orbit. 
As described in Section~\ref{ss:destabilization} for simulations near Neptune's 20:1 resonance, we ran a set of simulations for the 2:1 through 28:1 resonances in which the Sun and all four giant planets were fully interacting on their present-day orbits, but the test particles did not feel any direct gravitational effects from Uranus.
The residence time mosaic for this set of simulations is shown in Figure~\ref{fig:4-planet-mosaic} (we labeled these as `limited Uranus' in the figures) immediately alongside the full simulation.
There is a remarkable similarity between the two scenarios, especially at and beyond Neptune's 20:1 resonance.
The intermediate N:1 resonances (between the 10:1 and 20:1) do appear stronger when Uranus does not directly perturb the test particles, and this is reflected in the resonance strength measurements shown in Figure~\ref{fig:strength-data-poster}; we also include the averaged resonance strengths from the limited Uranus simulations in Figure~\ref{fig:average-strengths}.
When we compare the resonance strengths in Figure~\ref{fig:strength-data-poster} from the simulations with Neptune on its current orbit without Uranus, with limited Uranus effects, and with the full influence of Uranus, we note only small differences in the strengths of Neptune's resonances out to the $\sim10$:1, at which point the full simulation resonances strengths begin to decrease significantly; the limited Uranus simulation resonance strengths maintain similarity to the no Uranus simulations a bit further out before dramatically decreasing at the $\sim15$:1.
It is clear that the variations in Neptune's orbit due to Uranus are responsible for most of the weakening in Neptune's distant resonances, though more direct perturbations from Uranus have some influence on the intermediate resonances.

To briefly summarize what we conclude from our numerical simulations:
\begin{itemize}
    \item In the absence of Uranus, Neptune's exterior mean motion resonances in the {$q=33$~au} scattering population would be significantly stronger and extend out to much larger period ratios. 
    \item At the present day Neptune to Uranus period ratio (N:U = 1.96), nearly all of the weakening in Neptune's most distant resonances (beyond the $\sim20$:1) in the scattering population can be attributed to the variations in Neptune's orbit induced by the near-resonance with Uranus (particularly the $a$ variations) rather than to Uranus's direct gravitational influence on the scattering population
    \item Neptune's intermediate resonances in the present-day solar system (the $\sim10$:1 through the $\sim19$:1 are slightly weakened by Uranus's direct gravitational influence. 
    \item All of Neptune's external resonances would be significantly weaker if Neptune and Uranus were closer to their mutual 2:1 resonance, essentially vanishing at N:U=2.
\end{itemize}
We discuss the implications of these points in Section~\ref{sec:discussion}.

\section{Discussion} \label{sec:discussion}

Our exploration of Neptune's resonance strengths in the current giant planet configuration and alternative configurations has several important implications for the current and past evolution of the TNO populations.
First, we now have an explanation for the drop-off in resonance sticking at large $a$ in the scattering TNO population (seen by, e.g., \citealt{Lykawka:2007} in numerical models of the scattering population). 
The drop-off occurs because variations in Neptune's orbit (particularly its semimajor axis) induced by the near 2:1 resonance with Uranus essentially destabilizes Neptune's resonances beyond $a\sim200-250$~au entirely for nearly Neptune-crossing eccentricities ($q=33$~au).
If Neptune had migrated further outward by several au in semimajor axis in the early solar system and was thus better dynamically separated from Uranus, the present-day scattering TNO population would experience more resonance sticking at large semimajor axes. 
The variations in Neptune's orbit also slightly weaken Neptune's closer-in resonances, though there is additional weakening from Uranus's direct gravitational influence on the scattering population.

The simplest explanation for this additional destabilization of Neptune's present-day resonances by Uranus is overlap between Neptune's external resonances and Uranus's external resonances. 
This likely explains the rapid deterioration of stable libration in Neptune's resonances for the simulations in which we placed Neptune and Uranus at their exact 2:1 period ratio.
When Neptune and Uranus are in their mutual 2:1 resonance,
each of Neptune's exterior N:1 and N:2 resonances will be overlapped by an exterior N:1 resonance of Uranus. 
This overlap between multiple strong resonances causes the strengths of Neptune's N:1 resonances to drop off very quickly in Figure~\ref{fig:strength-data-poster}; in Figure~\ref{fig:4-planet-mosaic} there is no apparent stable libration in any resonance with Neptune beyond the $\sim5:1$ for the simulation with N:U = 2.
When Neptune and Uranus are not at their exact 2:1 period ratio, there will still be places where their exterior resonances happen to overlap or otherwise interact. 
\cite{Gallardo:2020} discusses how Uranus's resonances are faintly visible in dynamical maps of Neptune's resonances (they highlight Uranus's 4:1 resonance just exterior to Neptune's 2:1 in their Figure 23).
They note that Uranus's resonances are stronger at higher eccentricities, such as those in the low-$q$ scattering population we consider here (see also Figure 4 in \citealt{Gallardo:2018}). 
This is likely why Neptune's distant resonances regain strength at large $a$ for the lower-$e$, higher-$q$ orbits in the detached population (see Figure~\ref{fig:a-e-zoomed} and \citealt{Volk:2022}).

Overlap between Neptune's external resonances and three-body resonances involving Neptune, Uranus, and TNOs also potentially contribute to the weakening of Neptune's resonances.
Many main belt asteroids are known to exist in three-body resonances, mostly involving Jupiter and Saturn \citep[e.g.][]{Gallardo:2018}. 
Given the short orbital periods of main belt asteroids, their orbital parameters are typically much more tightly constrained than TNOs; it is thus more rare for TNOs to have orbit-fits accurate enough for secure identification in resonances as narrow as three-body resonances tend to be. 
\cite{Gallardo:2018} discuss 3 TNOs likely in the the 2 +1U-3N three-body resonance with Neptune and Uranus (a resonance identified by \citealt{Smirnov:2017}); they conclude that the influence of three-body resonances on Neptune's two-body resonances leads to chaotic diffusion but is not a problem that has been sufficiently well studied. 
Overlap between Neptune's external two-body resonances, Uranus's external two-body resonances, and potential three-body resonances with both planets likely plays a role in setting some of the smaller scale, noisier structures seen in the patterns of resonance strengths in {Figures~\ref{fig:strengths} and~\ref{fig:strength-data-poster}}.

Beyond resonance sticking in the present-day scattering population, the strong dependence of Neptune's distant resonance strengths on the Neptune to Uranus period ratio (and on Neptune's orbital variations due to interactions with Uranus) has very important implications for using Neptune's current resonant populations to constrain the migration history of the giant planets.
As the orbits of both Neptune and Uranus evolved during the epoch of giant planet migration, their period ratio and amount of dynamical coupling would vary over time. 
Thus the strengths of Neptune's distant external resonances could vary significantly depending on the exact interactions between Uranus and Neptune; this could dramatically change the likelihood of capturing TNOs into Neptune's resonances at different times during migration.
In particular, overall predictions for distant resonance populations from planet migration simulations could be very sensitive to how close Neptune and Uranus approach their mutual 2:1 resonance, with distant resonance capture probabilities likely dropping to near zero if the two planets spend significant time too close to that resonance.
{A recent exploration of more realistic planet migration modeling by \cite{Kaib:2024} found many simulations in which Neptune and Uranus crossed their mutual 2:1; they note that this crossing is unlikely to have happened in the real solar system, but these simulations highlight the expectation that the N:U period ratio during migration can be quite variable.}

The observed semimajor axis distribution of TNOs that are stably in Neptune's distant resonances and those on stable orbits that are near but \textit{not} in resonance provide powerful constraints on the migration history of the planets. 
The stable distant resonant TNOs are likely scattering TNOs that stuck to the resonances while Neptune was still migrating and were pushed into more stable phase space.
The strength of the resonances during migration, and thus the probability of sticking to them and being captured, varies depending on the N:U period ratio (Figure~\ref{fig:average-strengths}).
Thus the relative populations of captured TNOs across a wide range of Neptune's distant resonances will be a sensitive probe of the period ratio history of the two planets, though the observed populations are a combination of stable and transiently resonant objects \citep[e.g.][]{Crompvoets:2022} that must be disentangled.

The non-resonant TNOs on stable (i.e., not scattering orbits) sunward of these distant resonances are often interpreted as resonant drop-outs.
These could be TNOs that stuck to the resonances and evolved through secular interactions to low-enough eccentricities to be left behind as Neptune migrated \citep[e.g.][]{Gomes:2005}.
They also could have been dropped from resonance by discrete jumps in Neptune's semimajor axis during migration as Neptune interacted with the most massive planetestimals in the pre-migration disk \citep[e.g.][]{Nesvorny:2016,Kaib:2016,Lawler:2019}.
The sensitivity of Neptune's resonance strengths to the N:U period ratio hints at an additional possibility: resonance drop-outs could also occur if the two planets approach too closely to their mutual 2:1 resonance during migration.
While our investigation only focused on the low-$q$ population, and Neptune's distant resonances at the present-day N:U period ratio remain strong for the higher-$q$ orbits associated with the resonance drop-outs, the destabilization of Neptune's resonances is very dramatic as the planets approach their mutual resonance.
{To confirm that a mutual Neptune-Uranus resonance affects resonance strengths for the lower-$e$/high-$q$ TNO population, we re-ran our simulation with the N:U$=2$ period ratio with test particles distributed across Neptune's external resonances at $q=40$~au. 
Manual inspection of these simulation results reveals no apparent libration in any of Neptune's 2:1 through 28:1 through external resonances.
Resonance dropouts at higher-$q$ could indeed result from temporary excursions toward N:U$=2$ period ratios during migration.}

{It is clear that the distribution of TNOs in and near Neptune's distant resonances should be sensitive to the history of the N:U period ratio during migration. 
We note that the overall evolution of the scattering population could be sensitive to this as well.
For low-$q$ scattering objects, sticking to Neptune's resonances essentially `pauses' their diffusion in semimajor axis, increasing the time it takes for a particle to be removed from the scattering population. 
Temporary enhancements or reductions in the sticking strengths of Neptune's resonances during migration could change the decay rate of the overall scattering population. 
The history of the N:U period ratio could thus also affect how much of the primordial scattering population is retained during migration.}

\section{Summary and Conclusions}\label{sec:summary}

We have explored the phase space of Neptune’s mean motion resonances from $a\approx50-285$ au for eccentricities typical of scattering TNOs, finding:
\begin{itemize}
\item In the three-body problem (Sun, Neptune, and test particles) as well as in a solar system with Uranus removed (Sun, Jupiter, Saturn, Neptune, and test particles) Neptune’s external mean motion resonances would remain strong out to several hundred au in the scattering TNO population.
\item In the current solar system (Sun, Jupiter, Saturn, Uranus, Neptune, and test particles), Neptune’s resonances become unstable beyond $a\approx200-250$~au for low-$q$ orbits, explaining the lack of resonance sticking seen in simulations of the high-$a$ scattering TNO population.
\item The destabilization of Neptune's most distant resonances in the scattering population is due to variations in Neptune's orbit that are caused by the proximity of Uranus and Neptune to their mutual 2:1 resonance.
\item Uranus has a more direct influence on Neptune's closer-in resonances, weakening them sightly compared to simulations in which Uranus is not present; this is likely the result of overlap between Neptune and Uranus's two-body resonances as well as potentially weaker three-body resonances.
\item The strengths of all of Neptune's external resonances are quite sensitive to Neptune's period ratio with Uranus; if Neptune and Uranus are placed at their 2:1 resonance, there would be almost no stable libration in Neptune's resonances.
\end{itemize}

Our results imply that the drop-off in resonance sticking in the high-$a$ scattering TNO population \citep[e.g.,][]{Lykawka:2007} is not a generic outcome of small body scattering dynamics, but is instead a result of Uranus's perturbations on Neptune. 
The exact pattern of destabilization for Neptune's resonances as a result of Neptune-Uranus interactions is complex and has important implications for the behavior of Neptune's resonances during the era of planet migration when interactions between Neptune and Uranus were evolving. 
The overall number of stable distant resonant TNOs captured during migration, the relative stable populations between the distant resonances, as well as the number and distribution of resonant dropouts could be very sensitive to the history of Neptune and Uranus's period ratio.
Future, large-scale surveys such as the Vera Rubin Observatory's Legacy Survey of Space and Time (LSST) will dramatically improve observational constraints on the distant resonant, scattering, and detached TNO populations \citep[e.g.][]{Ivezic:2019}.
These observations should reveal complex patterns in the number of temporarily resonant scattering TNOs that reflect the relative strengths of the resonances.
The revealed distributions of stably resonant and detached TNOs will provide important constraints on the relative orbital histories of Uranus and Neptune.

\vspace{12pt}
\noindent{\it Acknowledgements:}
{We thank two anonymous referees for comments and suggestions that improved the manuscript.}
This work was supported by NASA (grant 80NSSC19K0785). 
KV acknowledges additional support from NASA (grant 80NSSC21K0376) and NSF (AST-1824869).
This work used High Performance Computing (HPC) resources supported by the University of Arizona TRIF, UITS, and Research, Innovation, and Impact (RII) and maintained by the UArizona Research Technologies department.

\facilities{ADS}
\software{rebound}

\appendix
\section{Resonance strength calculation}\label{sec:appendix}

As was explained in Section~\ref{ss:strengths}, we wanted to better quantify the strength of resonances that accounts for test particles that `stick' transiently in addition to stably librating particles.
We do this by considering total resonance occupation timescales as a fraction of simulation time.
For example, a test particle may start as non-resonant then move into Neptune’s 20:1 resonance, librate for 200 {orbits}, then be scattered away.
{If the duration of the simulation is 1000 20:1 orbital periods ($\sim$3.3 Myr)}, then that test particle can be said to be resonant for 20\% of the simulation time.
For this particular resonance, this one particle can be assigned a strength value of 0.2.
As discussed in Section~\ref{ss:strengths}, the nature of the sticking events and resonance behavior we are interested in makes it difficult to rely solely on the confinement of $\psi$ to determine periods of resonant interactions; the symmetric libration zones of Neptune's N:1 resonances in particular span nearly the entire $\psi=0-360^\circ$ range, and even particles in the the more confined resonant islands will experience slips in $\psi$ through the full range.
So instead we rely on relative confinement of semimajor axis to determine periods of resonant interactions.
We note that this method of determining resonant interactions only works for the low-$q$ scattering particles we are considering.
High-$q$ particles can remain at constant $a$ over significant timescales without any resonant interactions, whereas low-$q$ particles are very mobile in $a$ if they are not interacting with resonances.
Even if low-$q$ particles stay at constant $a$ without true resonant libration, they typically only do so because they are interacting with a resonance boundary, so it is reasonable to include them in the strength calculation.

We extend our analysis of individual test particle resonance interaction timespans to assign an overall strength value to a given resonance based on the entire set of test particles using the following steps:
\begin{itemize}
    \item Define a window in period ratio with Neptune around a target resonance.
    We define the boundaries of N:1 resonances with a window size of $\pm0.15$ in period ration with Neptune on either side of the exact resonant ratio; for N:2 resonances we use a period ratio window of  $\pm0.1$, and $\pm0.075$ for N:3 resonances.
    An example of this for the 20:1 resonance would be a period ratio with Neptune from 19.85 to 20.15, corresponding to $a=220.36-222.6$.
    These values adequately cover the expected width of each resonance type, without overlapping onto neighboring strong resonances.
    We only limit our target windows by period ratio, not $\psi$.
    \item Identify every test particle that passes through this window throughout the duration of the simulation and assign that particle a `residence time ratio', $s_i$, that describes the resonant interaction with a target resonance. This is done as follows:
    \begin{itemize}
        \item Define a sliding timescale window ($\Delta t$).
        We set this to {be 100 perihelion passages, which} is $\sim10$\% of the overall simulation time ($T=1000$ {orbital periods of the N:1 resonance nearest that particle's starting semimajor axis}). 
        For each particle that passes though our period ratio region of interest, we start the sliding time window at $t = 0$ and find the maximum and minimum period ratio recorded for that particle over $\Delta t$.
        If these values are wholly within the period ratio bounds, then we consider this particle `bound' to that phase space window in that given timespan of the simulation.
        \item The start of the sliding time window is then set to $t= 1$, then $t = 2$, then $t = 3$ {(the times of the next three perihelion passages)}, and so on until {the window end reaches the end of the simulation}, with the above determination made for each window.
        \item Once we have recorded the number $N_{res}$ of {all time windows} in which the test particle was bound to the period ratio range, we calculate the fraction of the total simulation time test particle $i$ was resonant: $s_i = N_{res}/(T- \Delta t)$.
        \item {Note: Our simulations output the orbital elements of each test particle at every perihelion passage. The timing between each perihelion passage is not always the same because the test particles are evolving in semimajor axis and thus not every particle achieves exactly 1000 perihelion passages over the simulation. However, when  a particle as resonant, its orbital period is the same as that of the target resonance, so the normalization of $s_i$ as a fraction of the expected 1000 resonant orbital periods is unaffected.}
    \end{itemize}
    \item A unitless strength value for a particular resonance can then be determined by the sum of residence time ratios ($s_i$) for all the particles that enter that region of phase space throughout the duration of the simulation.
    This is then normalized to the total number of test particles that are initialized in that region at the beginning of the simulation time ($n_{tp,0}$). The strength value $S$ is thus: 
    \begin{equation}
    S =  \frac{ \sum_0^N{s_{i}} }{ n_{tp,0} }. 
    \end{equation}
    We note that some of the particles whose $s_i$ values are included in $S$ were initialized outside of the period ratio range from which $n_{tp,0}$ is determined. 
    But because we are considering a low-$q$ scattering population, and many of the particle initialized near a resonance will scatter fairly quickly, the value of $S$ in our simulations is always between 0 and 1, and $S$ is approximately the average of all residence time ratios of for particles in the target period ratio window.
    This normalization scheme is tailored to our scheme of initializing particles along lines of constant $\psi$ and equally spaced in the semimajor axis; other normalizations would need to be considered for alternative test particle distributions.
\end{itemize}

\noindent As noted above, our strength parameter $S$ is best suited for the low-$q$ scattering population we focus on. 
It is also best used as a relative rather than absolute strength measurement; this is why we typically normalize resonance strengths to the value of $S$ for the strongest resonance in a simulation (e.g., Figure~\ref{fig:strengths}).
{There are many different ways resonance strength or stickiness can be parameterized. 
Our resonance strength parameter differs from the relative resonance stickiness parameter used by \cite{Lykawka:2007}; they normalize the residence time in individual resonances to the total residence time across all resonances, whereas we normalize to the total simulation time.
The overall trends in both versions of the resonance strength/stickiness are quite similar (see their Figure 5).
However, note that their simulations differ significantly from ours because they examine the entire scattering population over much longer timescales with far fewer test particles in each resonance due to the longer timescales.}

When plotting $a$-$\psi$ phase diagrams in Figure~\ref{fig:4-planet-mosaic}, we use a  modified version of a particle residence time to better highlight general particle stability.
We used the same sliding window method as described above for examining particle behavior for the duration of the simulation, but looked only at semimajor axis stability in each window rather than confinement to a specific period ratio range.
If the semimajor axis of a particle changed by no more than 2 au (roughly twice the width of typical resonances in the range we are interested in) we considered it stable for that time window. 
The fraction of all examined time windows for which a particle was stable is then the `normalized residence time' parameter used to color-code particles in the $a$-$\psi$ maps in Figure~\ref{fig:4-planet-mosaic}.
We found this scheme the most effective for visually illustrating the changes in resonant behavior in our different simulations.

\end{document}